\documentclass[11pt]{article}

\usepackage[a4paper, total={6in, 8.6in}]{geometry}

\usepackage{times}
\usepackage{color}
\usepackage{amsfonts}
\usepackage{graphicx}
\usepackage{amsmath,amsfonts,amssymb,bm}
\usepackage{booktabs} 
\usepackage{array}
\usepackage{boxedminipage}
\usepackage{paralist}
\usepackage{url}
\usepackage{diagbox}
\usepackage{enumitem}
\usepackage{commath}
\usepackage{paralist}
\definecolor{darkspringgreen}{rgb}{0.05, 0.5, 0.06}
\usepackage{arydshln}
\definecolor{Gray}{gray}{0.8}

\newcommand{\blue}[1]{\textcolor{blue}{#1}}

\newcommand{\sys}{\mbox{\sc NN-Crf}}
\newcolumntype{L}[1]{>{\centering\let\newline\\\arraybackslash\hspace{0pt}}m{#1}}
\newcolumntype{C}[1]{>{\centering\let\newline\\\arraybackslash\hspace{0pt}}m{#1}}
\newcolumntype{R}[1]{>{\centering\let\newline\\\arraybackslash\hspace{0pt}}m{#1}}

\begin{document}
\author{Yi Luan
\\
University of Washington\\
   {luanyi@uw.edu}}
\title{Information Extraction from Scientific Literature for Method Recommendation}
\date{}
\maketitle
\begin{abstract}
As a research community grows, more and more papers are published each year. As a result there is increasing demand for improved methods for finding relevant papers, automatically understanding the key ideas and recommending potential methods for a target problem. Despite advances in search engines, it is still hard to identify new technologies according to a researcher's need. Due to the large variety of domains and extremely limited annotated resources, there has been relatively little work on leveraging natural language processing in scientific recommendation.
In this proposal, we aim at making scientific recommendations by  extracting scientific terms from a large collection of scientific papers and  organizing the terms into a knowledge graph.   In preliminary work, we trained a  scientific term extractor using a small amount of annotated data and obtained state-of-the-art  performance by leveraging large amount of unannotated papers through applying multiple semi-supervised approaches. We propose to construct a knowledge graph in a way that can make minimal use of hand annotated data, using only the extracted terms, unsupervised relational signals such as co-occurrence, and structural external resources such as Wikipedia. Latent relations between scientific terms can be learned from the graph.  Recommendations will be made through graph inference for both observed and unobserved relational pairs. 
\end{abstract}
\section{Introduction}
 New technologies  are always built on previous discoveries and a great technological breakthrough can  stimulate a chain of developments across different scientific fields. 
As a research community grows, more and more papers are published each year which requires more and more human effort to read and understand.  Typical  information  researchers  seek when reading are the task of the  paper  and the methods to solve the task. On top of that,  researchers need to figure out how the paper  is related to their own research. While  there  are  review  papers  for  some  areas,  it is generally difficult to be comprehensive. It is even harder to get  knowledge of publications from all other related fields and choose the ones that can shed light on the researcher's problem.  Therefore, an intelligant way of recommending useful and relevant scientific information is in great demand.
 
 Current efforts to make scientific recommendation are limited to search engines such as Google Scholar,\footnote{\url{https://scholar.google.com/}} or Semantic Scholar.\footnote{\url{https://www.semanticscholar.org/}}
Even though the search engines can provide a way of obtaining all publications related to a certain search query, it is still hard to filter out or organize key information from the massive search engine output.

In order to make scientific recommendations, we need to extract and organize useful scientific knowledge in a structural way. One way of representing knowledge in a large text collection is to build a knowledge graph, which models information in the form of entities
and relationships between them~\cite{luan2018uwnlp,luan2018multi}. A knowledge graph is a representation of knowledge first used by Google to enhance its search engine's search results with rich semantic information gathered from a wide variety of resources.\footnote{\url{https://googleblog.blogspot.com/2010/07/deeper-understanding-with-metaweb.html}} The goal is that users would be able to get an answer to  their query without having to navigate to the sites and read the text themselves. 
A knowledge graph is an important step to transform text-based search results into semantically aware question answering services~\cite{hixon2015learning, waltz1978english}.  
Knowledge graphs are also used in several specialized
domains. For instance, Bio2RDF \cite{belleau2008bio2rdf} and Neurocommons \cite{ruttenberg2009life} are knowledge graphs that integrate multiple sources of biomedical information. These knowledge graphs have been used for drug discovery and treatment recommendation \cite{toutanova2016compositional,quirk2016distant}, which is very similar to our problem.

Once a scientific knowledge graph is constructed, the system can learn latent relation patterns from the observed scientific terms and generalize the patterns to predict unobserved scientific pairs. The system can retrieve observed information as well as predicting  the possible application of a new algorithm (or possible methods to solve a new task), and make recommendations based on the user query. For example, the Generative Adversarial Network (GAN) has been first introduced in machine learning communities in 2014, reaches its peaks in 2015, yet is not been applied  in NLP field until 2017. NLP researchers may want to know the possible NLP applications of GAN, which could be inferred from relational learning of the scientific knowledge graph.  

Since not much research has been done on scientific knowledge graph construction, annotated data for both scientific term and relation extraction is very limited, which becomes the main challenge to construct a knowledge graph in scientific domain. Previous work on low resource knowledge base construction includes open domain relation extraction (e.g. OpenIE~\cite{etzioni2008open}), which uses processed text strings between the two entities as relations and results in a completely unconstrained knowledge graph. Since the relations are not specified, there is no generalization of these relations and thus it is difficult to use in other systems. Another way to construct a knowledge graph is through distant learning, which requires a high precision, high coverage database that can be used for automatic annotation. However, few such database is available in scientific domain. Therefore we need to explore new approaches to construct the graph. In this thesis, we propose to apply  semi-supervised approaches to scientific information extraction  and  develop unsupervised approaches for knowledge graph construction and scientific recommendation. We will discover intrinsic relation signals embedded in the text such as co-occurrence and build the knowledge graph by leveraging both large scale unannotated data and structural external resources such as Wikipedia.  Experiments will be based on a large collection of papers from a wide coverage of AI communities which includes \textit{speech}, \textit{computer vision} and \textit{natural language processing}. 


The remainder of this proposal is structured as follows. In Section 2,  previous work on information extraction and knowledge graph construction is  reviewed. In Section 3,  our work on scientific term extraction using a semi-supervised neural approach is introduced. Section 4 introduces the dataset we use, and proposes the research plan for scientific knowledge graph construction and method recommendation. Conclusions are summarized in Section 5.
\section{Background}

The major task of Information Extraction (IE) is to turn unstructured text into structured information. Usually IE can be regarded as a pipeline of process with several different types of information that can be extracted: keyphrases (or, entity extraction) and relations between the keyphrases. 

\subsection{Entity Extraction}
A representative task in Entity Extraction is Named Entity Recognition (NER). It is usually modeled as a sequence tagging problem. In previous studies of NER, carefully constructed orthographic features and language-specific knowledge resources such as gazetteers are widely used~\cite{collobert2011natural,saha2008gazetteer,settles2004biomedical, luan2016lstm}. 
However, language-specific resources and features are costly to develop in new languages and new domains, making the approaches to NER a challenge to adapt. With the introduction of neural networks, the performance of NER systems has improved substantially~\cite{luan2015efficient,lample2016neural,peng2015named}. Neural approaches can replace hand-engineered features with well-designed structures that can be easily adapted to other domains or languages.

\subsubsection{Neural Sequential Tagging}
\label{sec:seq}
Sequence tagging has been a classic NLP task which includes part-of-speech tagging (POS), chunking, and named entity recognition (NER). 
Problems like chunking and NER require detecting the exact span of a term, which could include several tokens within a sentence. In order to  be able to distinguish spans of two consecutive terms of the same type, sentences are usually represented in the IOB format (Inside, Outside, Beginning) where every token is labeled as B-label if the token is the beginning of a named entity, I-label if it is inside a named entity but not the first token, or O otherwise.

Sequence tagging can be treated as a series of independent classification tasks, one per item of the sequence. However, accuracy is generally improved by making the optimal label for a given element dependent on the choices of nearby elements. Common models for sequence tagging are linear statistical models which include the Hidden Markov Model (HMM) and the Conditional Random Field (CRF) \cite{lafferty2001conditional}. With the introduction of neural approaches, the Convolutional Neural Network and Recurrent Neural Network (RNN) have been proposed to tackle the sequence tagging problem \cite{collobert2011natural,huang2015bidirectional}. Conventional RNNs with sigmoid units suffer from gradient decay or explosion, making training difficult. 
Long-Short-Term Memory (LSTM) \cite{hochreiter1997long}  models combat the vanishing gradient problem by using a series of gates to avoid amplifying or suppressing gradients during backpropagation. LSTMs have  proved to outperform other architectures in many NLP applications and have gained a lot of attention recently.

\begin{figure*}[t]
\centering
\includegraphics[width=7cm]{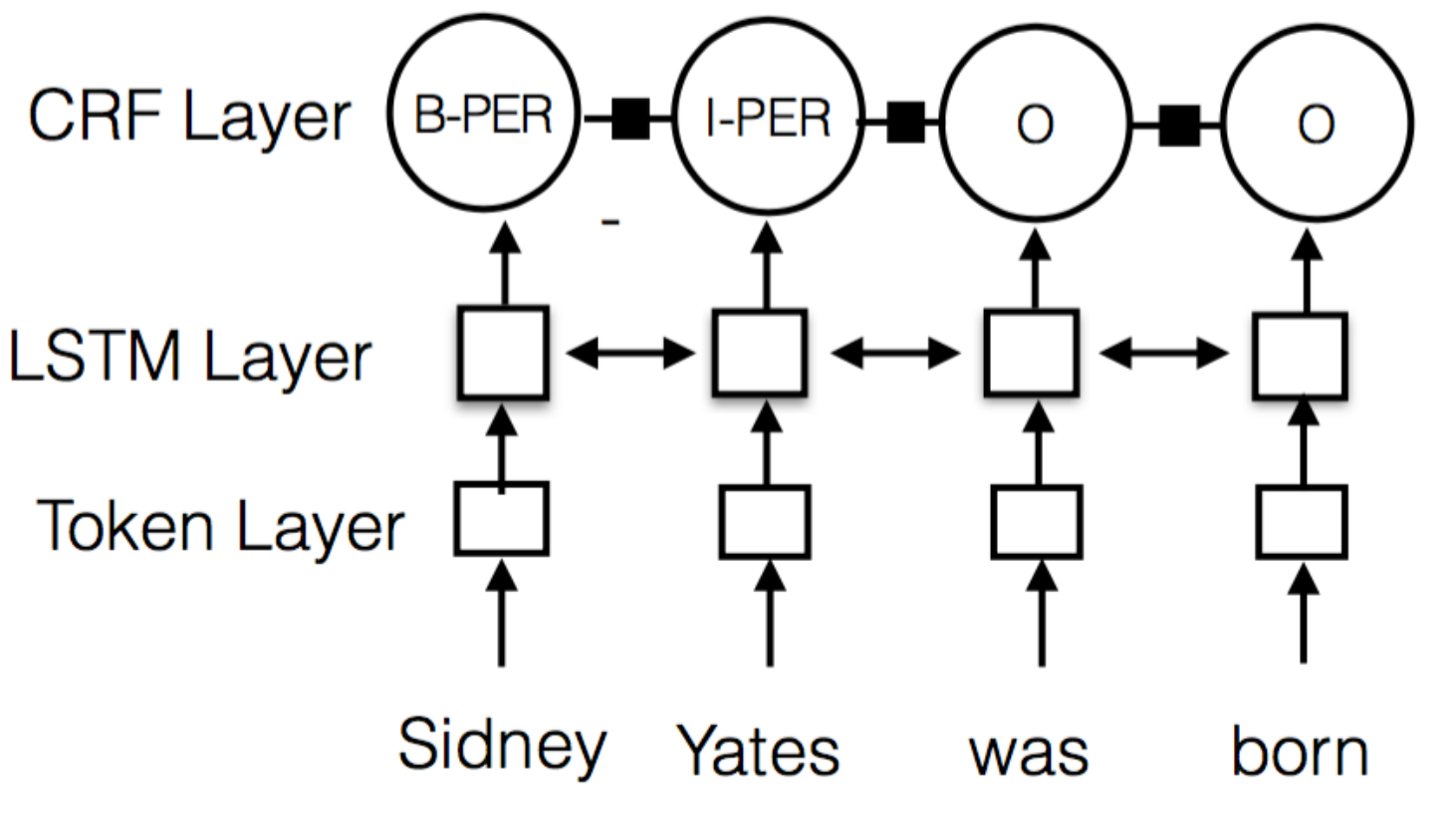}           
\caption{Neural Network Structure }
\label{fig:NN-NER}
\end{figure*}

There is also great interest in NNs that use character-based representations \cite{chiu2015named,ballesteros2015improved,ma2016end} to reduce the effect of out-of-vocabulary (OOV) words. 
Recent work \cite{lample2016neural} uses a CRF objective function on top of hybrid word-character LSTM structure and get state-of-the-art result in a NER task.
The sequence is tagged using a hierarchical multi-stage model that consists of 3 layers (Figure \ref{fig:NN-NER}): 
\begin{enumerate}
    \item The \textbf{Token Representation Layer} is the representation of each token, which can be a word embedding or a character representation.

    \item The \textbf{Token LSTM Layer} uses a bidirectional LSTM to incorporate contextual cues from surrounding tokens to derive intermediate token embeddings.
    
    \item The \textbf{CRF Layer} models token-level tagging decisions jointly using a CRF objective function.
\end{enumerate}

\paragraph{CRF Layer} 
For an input sentence $\bm{x}= (x_1,x_2,x_3,\dots,x_n)$, consider $M$ to be the matrix of scores output by
the bidirectional LSTM network. $M$ is of size $n \times m$,
where $n$ is the number of tokens in a sentence, and $m$ is the number of distinct tags.  $P_{i,t}$ corresponds to the score of the $i$-th tag of the $t$-th word
in a sentence: 
\begin{equation}
\label{eq:p}
P_{t,i}=p(y_t=i|x_t)
\end{equation}
A first-order Markov Model is used and a transition matrix $\bm{T}$ is defined where $\bm{T}_{i,j}$ represents the score from tag $i$ to tag $j$. $y_0$ and $y_n$ are added as the \textit{start} and \textit{end} tags of a sentence. Therefore $\bm{T}$ becomes a square matrix of dimension $m+2$. 

\noindent Given one possible output $\bm{y}= (y_1,y_2,y_3,\dots,y_n)$, and neural network parameters $\theta$ the score is defined as
\begin{equation}
\label{eq:fi}
\phi(\bm{y};\bm{x},\theta) = \sum_{t=0}^n T_{y_t,y_{t+1}} + \sum_{t=1}^n P_{t,y_t}
\end{equation}
The probability of sequence $\bm{y}$ is obtained by doing a softmax
\begin{equation}
\label{eq:score}
p_\theta(\bm{y}|\bm{x}) =  \frac{\exp(\phi(\bm{y};\bm{x},\theta))}{\sum_{\bm{y}'\in \bm{y}^m} \exp(\phi(\bm{y}';\bm{x},\theta))}
\end{equation}

\noindent where $\bm{y}^m$ is the space of all possible tag sequences. The forward algorithm can be used to efficiently calculate the denominator.

\subsection{Relation Extraction}
Relation Extraction is the next step in analyzing information in texts and turning unstructured information into structured information.  After the information is structured and added to a database, it can be used by a wide range of NLP applications, including information retrieval, question answering and many others. Therefore, Relation Extraction is a very important step to knowledge base completion. There are two types of methods used for relation extraction: self-supervised methods and supervised methods.

For scenarios with no labeled data but large amounts of unlabeled data and a small set of extraction patterns, self-supervised systems make the process of relation extraction largely unsupervised.
The KnowItAll Web IE system \cite{etzioni2005unsupervised} is an
example of a self-supervised system.
In order to make IE systems faster and more scalable, \cite{banko2007open,etzioni2008open} introduces \textit{Open IE}, which does not presuppose a predefined set of relations and is targeted at all relations that can be extracted. The system makes a single data-driven pass over its corpus and extracts a large set of relational triplets without requiring any human input.  The output of this system consists of triplets stating there is some relation between two entities, but since the relations are not specified, they are difficult to  use  in some other systems.

Supervised methods rely on a training set where domain-specific examples have been tagged. A pre-defined relation is  in the form of a triplets $t=(e_x,r_{xy},e_y)$ where $e_x,e_y$ are entities in a predefined relation $r$ within document $D$. For example, \textit{``Mike lives in Chicago."}, a relation \textit{(SMike, PHYS, Chicago)} would be extracted where \textit{PHYS} indicates \textit{located at}.
Such systems automatically learn extractors for relations by using machine-learning techniques. The main problem of using these methods is that the development of a suitably tagged corpus can take a lot of time and effort.

\subsection{Knowledge Base Representation Learning}
Knowledge base (KB) store collections of relation triples $t=(e_x,r_{xy},e_y)$ from relation extractor. Even the largest of knowledge bases are incomplete which motivate research of predicting missing information in knowledge bases. In order to recover missing triples, a statistical model needs to exploit regularities in the graphs.  Consider for example, the triple (\textit{RNN}, \textit{is\_a\_method\_of}, \textit{Sequential Tagging}) and (\textit{Named Entity Recognition}, \textit{is\_a\_task\_of}, \textit{Sequential Tagging}), then we can infer that \textit{Named Entity Recognition} can be solved by the method of \textit{Recurrent Neural Network}. The task of predicting missing relations are usually called Link Prediction, which is a subtask of Knowledge Base Completion.

In order to infer unobserved relations, knowledge base embedding approach was first proposed as an alternative statistical relational learning method, and has gained a significant amount of attention, due to its simple prediction time computation and strong empirical performance.   Early models are learned solely from known direct relationships between two entities (\textit{RNN},  \textit{is\_a\_method\_of},  \textit{Sequential Tagging}). In contrast, using multi-step relation paths (e.g. (\textit{NER}, \textit{is\_a\_task\_of}, \textit{Sequential Tagging})  $\wedge$ (\textit{RNN},  \textit{is\_a\_method\_of}, \textit{Sequential Tagging})) to train embeddings has been proposed and offered significant gains in embedding models for KB completion.

\subsubsection{Single-step KB learning}
\label{sec:singleKB}
In this framework, entities and relations in a knowledge base are represented in a continuous space. Whether two entities have a previously unknown relationship can be predicted by simple functions of their corresponding vectors \cite{yang2014embedding}, matrices \cite{nickel2011three} or tensors \cite{riedel2013relation}.

\paragraph{Vector Embedding}
In \cite{yang2014embedding}, $e_x$ and $e_y$ in relation triple $t=(e_x,r_{xy},e_y)$ are represented by embeddings $\bm{v}_{e_x}, \bm{v}_{e_y}$ and each relation type $r$ is represented by a matrix $\bm{Q}_r$.  Different  scoring functions are designed and compared, all which are defined by  a functions between entitiy pair embeddings  and relation matrix.
The  choice  of  relation  representations  are reflected  in  the  form  of  the  scoring  function for each triplet.  Most  of  the
existing scoring functions in the literature \cite{yang2014embedding, toutanova2016compositional} uses bilinear transformation as
\begin{equation}
\label{eq:bilinear}
S_{(e_x,r,e_y)} = \bm{v}_{e_x}^T \bm{Q}_r \bm{v}_{e_y}
\end{equation}

In order to encourage the scores of positive relationships (triplets)
to be higher than the scores of any negative relationships (triplets), the model is trained by  minimizing a margin-based ranking objective. Since only positive triplets
are observed in the data, given a set of positive triplets
$T$, a set of ``negative" triplets $T'$ is constructed by corrupting either one of the relation arguments: $T'=\{(e'_x, r, e_y)\not\in T\}\cup \{(e_x, r, e'_y)\not\in T\}$. The training objective is designed as minimizing the ranking loss:
\begin{equation}
\label{eq:objective}
L = \sum_{(e_x, r, e_y)\in T} \sum_{(e_x', r, e_y')\in T'} \max\{S_{(e_x',r,e_y')}- S_{(e_x,r,e_y)}+1,0\}
\end{equation}

Link prediction is then formulated as an entity ranking task. For each triplet in the test data, each entity is treated as the target entity to be predicted in turn. Scores are computed for all the entities in the dictionary and are ranked in descending order. The candidates with higher ranks are considered as correct link.

\paragraph{Matrix Factorization}
Another way of KB learning is to predict relation based on term occurrence and learn latent representation for terms and relations via matrix factorization.
Riedel et. al. \cite{riedel2013relation} generalize the surface forms of relations such as OpenIE \cite{etzioni2008open}, which serves as auxiliary relations for link prediction.

\begin{figure*}[t]
\centering
\includegraphics[width=8cm]{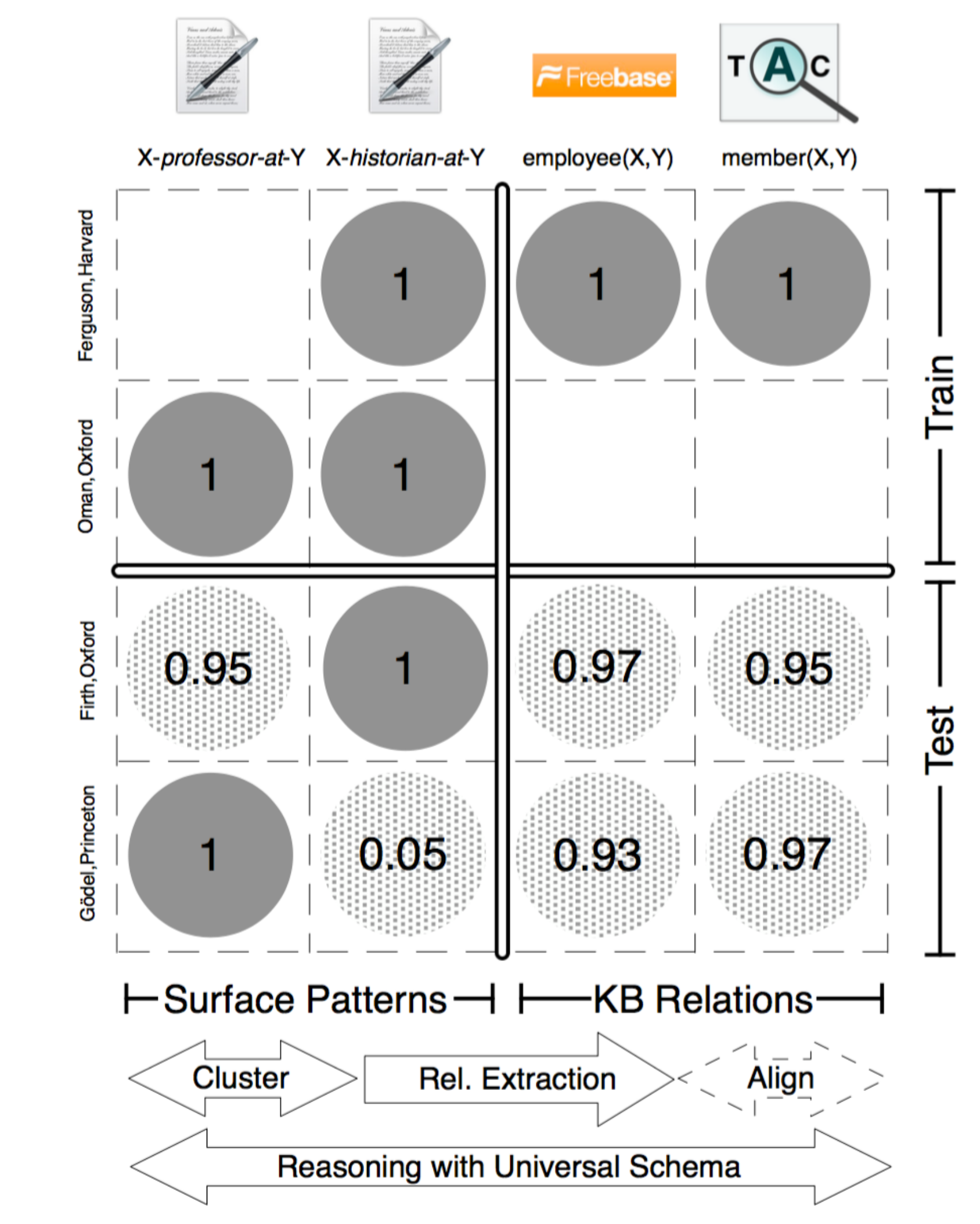}       
\caption{Example of relation extraction through matrix factorization, Figure from \cite{riedel2013relation}.
Dark circles are observed facts, shaded circles are inferred facts. Relation Extraction  maps surface pattern relations (and other features) to structured relations.
Surface form clustering models correlations between patterns.}
\label{fig:mat_fac}
\end{figure*}

In \cite{riedel2013relation}, the task is to learn latent feature vectors for entity triplets and relations,  columns correspond to relations, and rows correspond to entity triplets. An example of relation extraction through matrix factorization is as Fig.~\ref{fig:mat_fac}. The goal is a model that can estimate the probability of whether a missing element in the matrix is true.

A series of exponential family models that estimate this probability is introduced, which includes modeling three aspects: Latent Feature, Neighborhood Model and Entity Model. Latent Feature guarantees the low rank nature of the reconstructed matrix.  Neighborhood Model interpolates the confidence for a given triplet and Entity Model models latent entity representation and relation-argument representation. The scoring function for each triplet is to add all scores of all three models. 
Similar to Eq. \ref{eq:objective}, the objective function is to maximize a sum terms of all facts such that the score for each
observed fact is bigger than unobserved fact.

\begin{figure*}[t]
\centering
\includegraphics[width=8cm]{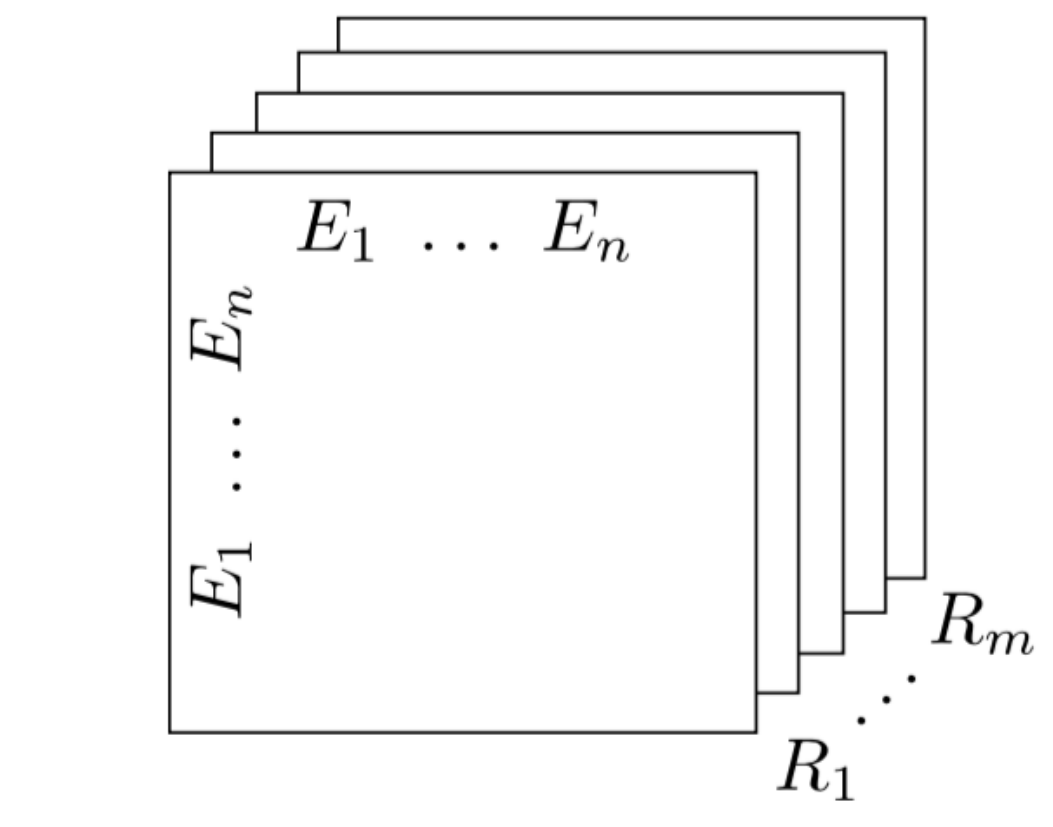}       
\caption{Tensor model for relational data.
$E_1\dots E_n$ denote the
entities, while $R_1\dots R_m$
denote the relations in the domain. Figure from \cite{nickel2011three}.}
\label{fig:mat_fac1}
\end{figure*}

\paragraph{Tensor based approaches}
Instead of clustering surface text patterns to relations, \cite{nickel2011three} focus on mining unobserved relations through existing ones via tensor. Given a number of entity $E_1\dots E_n$ and relations $R_1\dots R_m$, the goal is to infer missing relations for existing entities. A tensor $\mathcal{X}$ is constructed as in Fig.~\ref{fig:mat_fac1}, where each slice is a relation type and each cell in the slice indicates the relational existence of two entities. A  rank-$r$ factorization is employed , and each slice $X_k$ is factorized as
\begin{equation}
\mathcal{X}_k \approx AR_kA^T, \mathrm{for} \quad k = 1,\dots,m
\end{equation}
Here,
$A$ is a $n\times r$ matrix that contains the latent-component
representation of the entities in the domain and $R_k$
is an asymmetric $r\times r$
matrix that models the interactions of the latent  components in the $k$-th predicate. The model is optimized by a square loss between the reconstructed tensor with the original value.

\subsubsection{Multi-step KB learning}
\label{sec:multiKB}
Despite the success of single-step KB learning, it is known that there are also substantial multiple-step relation paths between entities indicating their semantic relationships. The relation paths reflect complicated inference patterns among relations in KBs. For example, the relation path (\textit{$e_1$}, \textit{$BornInCity$}, \textit{$e_2$})  $\wedge$ (\textit{$e_2$},  \textit{$CityInCountry$}, \textit{$e_3$}) indicates the relation $Nationality$ between $e_1$ and $e_3$.
Previous work about using multi-step relation paths for training can be roughly classified into two methods: (1) creating auxiliary triples and then add the triples to the learning objective of a factorization model \cite{guu2015traversing}; (2) using paths or walks as features when predicting edges \cite{lin2015modeling,toutanova2016compositional, schlichtkrull2017modeling}. 

Both approaches take into account embeddings of relation paths between entities, and both of them used vector space compositions to combine the embeddings of individual relation links $r_i$ into an embedding of the path $\pi=\{r_1\dots r_n\}$. The intermediate nodes $e_i$ are neglected. The composition function of a bilinear model is matrix multiplication. The embedding of a length-$n$ path $\Phi_\pi$ is defined as the matrix product of the sequence of relation matrices.
\begin{equation}
\label{eq:bilinear_multi}
    \phi_\pi = Q_{r_1}\dots Q_{r_n}
\end{equation}

In \cite{guu2015traversing}, information from relation paths was used to generate additional auxiliary terms in training, which provides more training triples for graph learning. The method starts from each node in the knowledge base, samples $m$ random walks of length 2 to a maximum length $L$, resulting in a list of triples $\{(e_x^{(i)},\pi^{(i)},e_y^{(i)})\}$. $e_x^{(i)}$ and $e_y^{(i)}$ are the start and end nodes of the random walk, and $\pi^{(i)}$ is the relation path between them. The score of each triple is defined as $f(e_x,\pi,e_y|\theta)=\bm{v}_{e_x}^T\Phi_\pi\bm{v}_{e_y}$. 

Instead of using relation paths to augment the set of training triples, \cite{lin2015modeling} proposed to used paths to define the scoring function $f(e_x,\pi,e_y|\theta,F(e_x,e_y))$. Here $F(e_x,e_y)$ denotes the sum of the embeddings of a set of path $\pi$ between the two nodes in the graph, weighted by path-constrained random walk probabilities as follows:
\begin{equation}
F(e_x,e_y)=\sum_{\pi} w_{|\pi|}P(e_y|e_x,\pi)\phi(\pi)
\end{equation}
$w_{|\pi|}$ is a shared parameter for path length, so that the model may learn to trust the contribution of different path lengths for different resources. $P(e_y|e_x,\pi)$ is the random walk probability for each path.

The score for each triplet is then defined as: 
\begin{equation}
\label{eq:score_multi}
f(e_x,\pi,e_y|\theta,F(e_x,e_y)) = \bm{v}_{e_x}^T \bm{Q}_r \bm{v}_{e_y} + \mathrm{vec}(F(e_x,e_y))^T\mathrm{vec}(\bm{Q}_r)
\end{equation}

The first term of the scoring function is the same as that of the bilinear model, and the second term takes into account the similarity of the weighted path representation and the predicted relation $r$.  
Element-wise product of two matrices is selected as the similarity metric. On top of this framework, \cite{toutanova2016compositional} propose a dynamic programming method that can learn all paths between two nodes without making any approximations like sampling or pruning, which achieved significant improvement over previous methods. The learning method is the same as the single-step, which is to minimize a ranking loss objective. 

\subsection{Research on Scientific Literature}
There has been growing interest in research on automatic methods to help researchers search and extract useful information from scientific literature. Past research on this field mainly focused on the following aspects. Some research investigated on citation function, for example analyzing citation sentiment, and predicting the reason for whether citing a paper is positive or negative \cite{athar2012detection,athar2012context}. Some research focused on citation network and community \cite{do2013extracting,jaidka2014computational}, where the main research problems are about exploring key authors in a field \cite{kas2011structures, sim2012discovering}, observation of over-time trends in these networks~\cite{vogel2012he, prabhakaran2016predicting}, and detecting scientific breakthroughs using text content~\cite{anderson2012towards}. Research on summarizing scientific papers has also been extensively explored~\cite{abu2011coherent}. However, due to scarce hand-annotated data resources, previous work on information extraction (IE) for scientific literature is very limited. Gupta and Manning \cite{gupta2011analyzing} first proposed a task that defines scientific terms into three aspects: \textit{domain}, \textit{technique} and \textit{focus} and apply template-based bootstrapping to tackle the problem. Based on this study, \cite{tsai2013concept} improve the performance by introducing hand-designed features from named entity recognition \cite{collins1999unsupervised} to the bootstrapping framework.

For relation extraction in scientific literature, most work has been done in the biomedical domain under a distant learning framework e.g. using  Gene  Drug  Knowledge  Database
 \cite{dienstmann2015database, quirk2016distant, peng2017cross}. The main challenge for relation extraction in scientific domains is the long context window that the relations can embed in.  Relations between scientific terms cross-sentence can be chained with coreference \cite{gerber2010beyond, yoshikawa2011coreference} and discourse relations \cite{quirk2016distant, peng2017cross}. However the performance of coreference and discourse parsers is not sufficiently accurate on scientific domains and not so much previous work has been researched.
There has been various proposed schema in scientific discourse analysis such as  \cite{teufel2002summarizing,de2012verb}, but mostly on limited hand annotated data. 

\subsection{Leveraging external resources}
The bottleneck of the supervised methods in information extraction is usually the lack of training data.
Therefore, leveraging large unlabeled text sources is very important. Previous work has mainly focused on transfer learning \cite{dai2015semi, luan2014semi}, multi-task learning \cite{collobert2008unified,peng2015named, luan2017multi} or initializing the model with pre-trained word embeddings \cite{mikolov2013efficient,pennington2014glove,levy2014dependency, luan2016multiplicative}. Here we especially focus on two ways of leveraging external resources: semi-supervised learning and distant learning. Semi-supervised learning (SSL) uses large scale unlabeled data to improve the performance of a supservised system, while distant learning is method to label unannotated data with the help of database.
\subsubsection{Semi-supervised Learning}
The earliest SSL algorithms use the Expecatation-Maximization (EM) algorithm. It assumes a model $p(x,y)=p(y)p(x|y)$ where $p(x|y)$ is an identifiable mixture distribution, for example Gaussian mixture models.  With large amount of unlabeled data,  the  mixture  components  can  be  identified;  only small amount of labeled data per component will be needed to  fully determine the  mixture distribution. EM based semi-supervised methods have been successfully applied to many applications such as text classification~\cite{nigam2000text} and face orientation  discrimination~\cite{baluja1998probabilistic}.

Another common SSL algorithm is self-training~\cite{scudder1965probability},
where one makes use of a previously trained model
to annotate unlabeled data which is then used to
re-train the model. Self-training  has  been successfully applied  to  several  natural  language  processing tasks such as word sense disambiguation \cite{nigam2000text}, parsing and machine translation~\cite{riloff2003learning}. 
The EM approach  can be viewed as a special case of soft self-training.  One can imagine that a classification mistake can reinforce itself. Some algorithms try to avoid this by ignoring unlabeled points if the prediction confidence is below a threshold.

Graph-based SSL algorithms \cite{zhu2003semi,corduneanu2002information,subramanya2009entropic}
are an important subclass of semi-supervised techniques that
have received much attention in the recent past.
Graph-based semi-supervised methods define a graph where the nodes are represented as the labeled and unlabeled samples in the dataset, and the edges are the similarity between the samples. These methods usually assume neighboring nodes on the graph tend to have similar output (smoothness over the graph).  Graph-based methods
are non-parametric, discriminative, and transductive in nature.
 Graph-based methods have also been widely used in NLP applications, but mostly focus on unstructured problems such as text classification~\cite{subramanya2008soft,ozaki2011using} and machine translation~\cite{alexandrescu2009graph}. Structured NLP problems such as POS tagging and NER tend to be hard to construct the graph, since it is hard to use  whole  sequence similarity to constrain whole tag sequences assigned
to  linked  examples. Subramanya et. al. \cite{subramanya2010efficient} construct the graph based on trigram similarity together with some hand-designed features and use graph-based semi-supervised on top of CRF structure to improve POS tagging performance. Following \cite{subramanya2010efficient}, similar methods has also been applied to NER~\cite{hakimov2012named} and slot filling tasks~\cite{aliannejadi2017graph}.

\subsubsection{Distant Learning}
Mintz et. al. \cite{mintz2009distant} introduce a new term ``distant supervision". The authors use a large semantic database Freebase containing 7,300 relations between 9 million named entities to annotate a large unlabeled corpus by matching the terms. A relation classifier is extracted on the annotated data. 
There are two main problems of distant learning: (1) some training examples obtained through this
heuristic are not valid, and (2) the same pair of entities can have
several relations. Many approaches to on handling the noisy annotation have been explored \cite{yao2010collective,yao2013universal,riedel2013relation, luan2014relating,levow2014recognition, luan2012performance}. Some researchers target the heuristics that are used to
map the relations in the databases to the texts, for example, \cite{takamatsu2012reducing} argue that improving matching helps to make data less noisy and therefore enhances the quality of relation extraction in general.
Despite the existing problems, distant learning has been applied widely in relation extraction, and provide sufficient training data. Many of the neural based systems are developed with the help of distant learning \cite{zeng2015distant,peng2017cross}.

\section{Initial work}
\label{sec:pilot}
In initial work, we address the problem of extracting  keyphrases from scientific articles and categorizing them as corresponding to a task, process, or material. We cast the problem as a sequence tagging problem and introduce  semi-supervised methods to a neural tagging model, which builds on recent advances in named entity recognition. Since annotated training data is scarce in this domain, we introduce a graph-based semi-supervised algorithm together  with a data selection scheme to leverage unannotated articles. Both inductive and transductive semi-supervised learning strategies outperform state-of-the-art information extraction performance on the 2017 SemEval Task 10 ScienceIE task. 
\begin{figure}
\centering
\begin{footnotesize}
\begin{tabular}{p{14cm}}
\toprule
{\it \textbf{Computer Science:}} \\ This paper addresses the tasks of \textbf{[Named Entity Recognition]$\blue{\mathrm{_{\bm{Task}}}}$}(\textbf{[NER]$\blue{\mathrm{_{\bm{Task}}}}$}), a subtask of \textbf{[information extraction]$\blue{\mathrm{_{\bm{Task}}}}$}, using  \textbf{[conditional random fields]$\blue{\mathrm{_{\bm{Process}}}}$}. Our method  is evaluated on the \textbf{[ConLL NER Corpus]$\blue{\mathrm{_{\bm{Material}}}}$}. 
\\
\midrule
{\it \textbf{Physics:}} \\ \textbf{[Local field effects] $\blue{\mathrm{_{\bm{Process}}}}$} on spontaneous emission rates within \textbf{[nanostructure photonics material]$\blue{\mathrm{_{\bm{Material}}}}$} for example are familiar, and have been well used.\\
\midrule
{\it \textbf{Material Science:}} \\ The \textbf{[Kelvin probe force microscopy technique] $\blue{\mathrm{_{\bm{Process}}}}$} allows \textbf{[detection of local EWF]$\blue{\mathrm{_{\bm{Task}}}}$} between an  \textbf{[atomic force micorscopy]$\blue{\mathrm{_{\bm{Material}}}}$} and  \textbf{[metal surface]$\blue{\mathrm{_{\bm{Material}}}}$}.\\
\bottomrule
\end{tabular}
\end{footnotesize}
\footnotesize{\caption{Annotated ScienceIE examples.}}
\label{fig:pp}
\end{figure}

\subsection{ScienceIE Dataset}
\label{sec:ScienceIE}
Very recently a new challenge on Scientific Information Extraction (SemEval Task 10 \textsc{ScienceIE})~\cite{scienceIE} provides a dataset consisting of 500 scientific paragraphs with keyphrase annotations  for three categories: \textsc{Task}, \textsc{Process}, \textsc{Material} across three scientific domains, Computer Science (CS), Material Science (MS), and Physics (Phy), as in Figure~\ref{fig:pp}. This dataset enables the use of more advanced approaches such as neural network (NN) models~\cite{scienceIE}. Identifying keyphrases is a very challenging subtask,
since the dataset contains many long and infrequent keyphrases.

In addition to keyphrase extraction, the ScienceIE also extracts semantic relations between keywords, including two types of relations: \textsc{HYPONYM-OF} and \textsc{SYNONYM-OF}. For example, in the first row of Figure~\ref{fig:pp}, \textit{Named Engity Recognition} is a \textsc{SYNONYM-OF} of \textit{NER}, while \textit{information extraction} is \textsc{HYPONYM-OF} of both \textit{Named Engity Recognition} and \textit{NER}. The relation extraction task of ScienceIE is not studied in initial work, but will be used for evaluating entity linking performance in Sec.~\ref{sec:linking}

The  ScienceIE  corpus consists of 500 journal articles; one paragraph of each article is randomly selected and annotated. The complete unlabeled articles and their metadata are provided together with the labeled data. The training data consists of 350 documents; 50 are kept for development and 100 for testing. The 500 articles come from 82 different journals evenly distributed in three domains. We manually labeled 82 journal names in the dataset into the three domains and do analysis based on the domain partitions.  The 500 full articles contains 2M words and is 30 times the size of the annotated data. 

\subsection{Neural Model and Semi-supervised Learning}
We adopt the 3-layer LSTM-CRF neural model \cite{lample2016neural} as in Fig.~\ref{fig:NN-NER} described in Sec.~\ref{sec:seq}  to tag scientific terms. 
We extend the token representation layer to be a concatenation of three components for each token:a bi-directional character-based embedding, a word embedding, and an embedding associated with orthographic and part-of-speech features. 

We develop a semi-supervised algorithm that extends self-training by estimating the labels of unlabeled data and then using those labels for re-training. Specifically, we use a  graph-based algorithm to estimate the posterior probabilities of unlabeled data 
and develop a new  CRF training to take the uncertainty of the estimated labels into account while optimizing the objective function.

\subsubsection{Graph-based Posterior Estimates}
\label{sec:graph}
Our semi-supervised algorithm uses the following steps to estimate the posterior. It first constructs a graph of tokens based on their semantic similarity, then uses the CRF marginal as a regularization term to do label propagation on the graph. The smoothed posterior is then used to either interpolate with the CRF marginal or as an additional feature to the neural network.

\paragraph{Graph Construction} Vertices in the graph correspond to tokens, and edges are distance between token features which capture semantic similarity. The total size of the graph is equal to the number of tokens in both labeled data $V_l$ and unlabeled data $V_u$. The tokens are modelled with a concatenation of pre-trained word embeddings (with dimension $d$) of 5-gram centered by the current token, the word embedding of the closest verb,  and a set of discrete features including part-of-speech tags and capitalization (43 and 4 dimension one-hot features). The resulting feature vector with dimension of $5d+d+43+4$ is  then projected down to 100 dimensions using PCA.   We define the weight $w_{uv}$ of the edge between nodes   $u$ and $v$ as follows: $w_{uv} = d_e(u,v) \text{ if } v \in \mathcal{K}(u) \text{ or } u\in \mathcal{K}(v)$,   where $\mathcal{K}(u)$ is the set of \textit{k}-nearest neighbors of $u$ and $d_e(u,v)$ is the Euclidean distance between any two nodes $u$ and $v$ in the graph.

For every node $i$ in the graph, we compute the marginal probabilities $\{ \bm{q}_i\}$ using the forward-backward algorithm. Let $\theta^i$ represent the estimate of the CRF parameters after the $n$-th iteration, we compute  the marginal probabilities 
$\tilde{\bm{p}}_{(j,t)} = p(y_t^{j}|\bm{x};\theta^i)$ over IOB tags for every token position $t$ in sentence $j$ in labeled and unlabeled data.

\paragraph{Label Propagation} 
We use prior-regularized measure propagation \cite{liu2014graph,subramanya2011semi} to propagate labels from the annotated data to their neighbors in the graph. The algorithm aims for the label distribution between neighboring nodes to be as similar to each other as possible by optimizing an objective function that
minimizes the Kullback-Leibler distances between: i) the empirical distribution $\bm{r}_u$  of labeled data and the predicted label distribution $\bm{q}_u$ for all labeled nodes in the graph; ii) the distributions $\bm{q}_u$ and $\bm{q}_v$ for all nodes $u$ in the graph and their neighbors $v$; iii) the distributions $\bm{q}_u$ and the CRF marginals $\tilde{\bm{p}}_u$ for all nodes. 
The third term regularizes the predicted distribution toward the CRF prediction if the node is not connected to a labeled vertex, ensuring the algorithm performs at least as well as standard self-training. 

\paragraph{Posterior Estimates}
We develop two strategies to estimate the new posteriors $\hat{p}(y_t|\bm{x};\theta)$, which can then be used in the CRF training.   

The first strategy (called \textsc{GraphInterp}) is the commonly used approach~\cite{subramanya2010efficient,aliannejadi2014graph}  that interpolates the smoothed posterior  $\{\bm{q}\}$ with CRF marginals $p$:
\begin{equation}
\label{eq:inductive}
\hat{p}(y_t|\bm{x};\theta) = \alpha p(y_t|\bm{x};\theta) + (1-\alpha) q(y)
\end{equation}
\noindent where $\alpha$ is a mixing coefficient. 

The second strategy (called \textsc{GraphFeat}) uses the smoothed posterior $\{\bm{q}\}$ as  features and learns it with other parameters in the neural network. 
Given a sentence $\{x_1,\dots,x_{n}\}$, let $Q=\{\bm{q}_1,\dots,\bm{q}_n\}$ be the predicted label distribution from the graph. We then use $Q$ as a feature input to neural network as 
 $\tilde{P} = P + M Q$
 where $P$ is the $n \times m$ matrix output by  the  bidirectional  LSTM network as in Eq. \ref{eq:fi}, and $M$ is $m\times m$ matrix and is learned together with other parameters of neural network.
 We modify Eq. \ref{eq:fi} by replacing ${P}_{t,y_t}$ with  $\tilde{P}_{t,y_t}$.
%
Note that \textsc{GraphFeat} can only be done in a transductive way  since it requires output $Q$ from the graph at test time.

\subsubsection{CRF training with Uncertain Labels}
\label{sec:ulm}

\begin{figure}[t]
\footnotesize
\centering
\includegraphics[width=8cm]{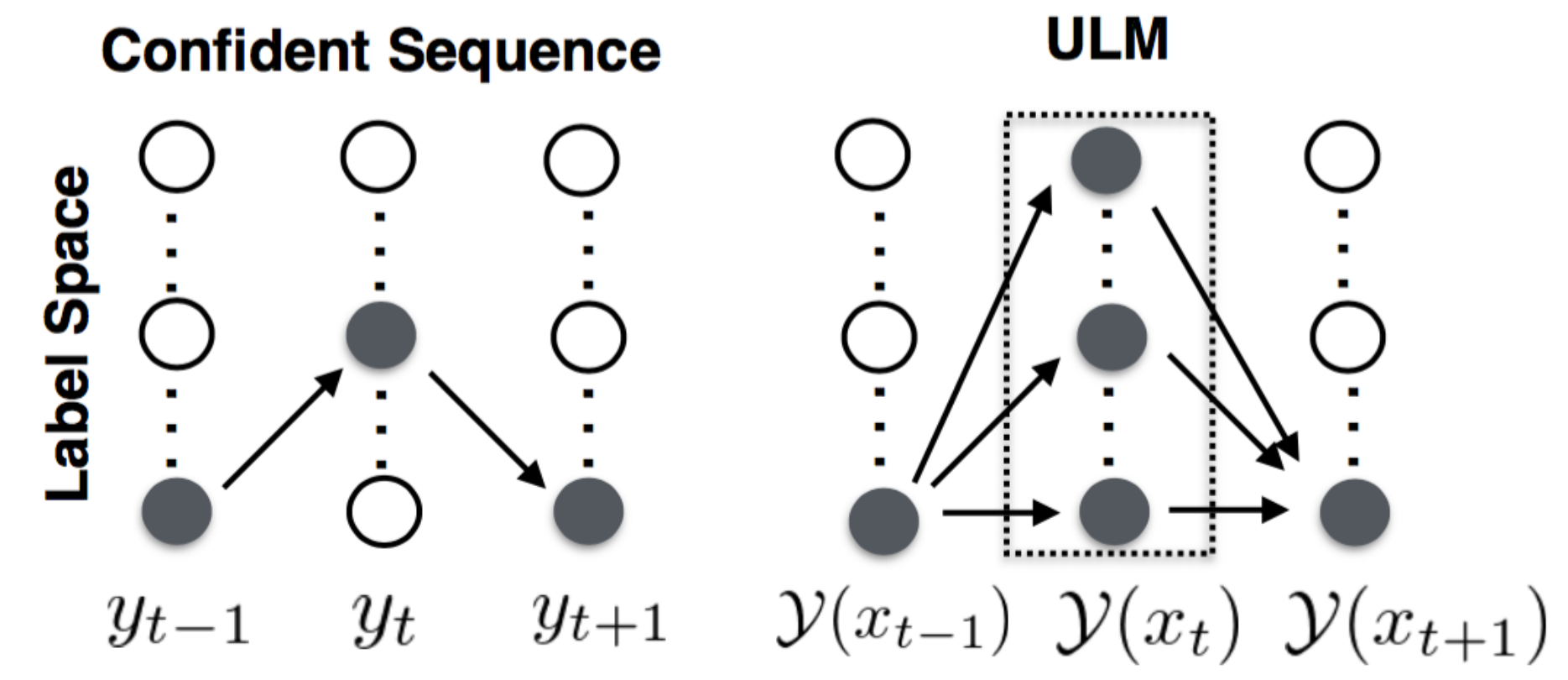}           
\caption{\small{Lattice representation of ULM. Dashed box is the uncertain token which is going to  be marginalized over.}}
\label{fig:ULM}
\end{figure}
A standard approach to self-training is to make hard decisions for labeling tokens based on the estimated posteriors and retrain the model.  However, the  
estimated posteriors in our task are noisy due to the difficulty and variety of the ScienceIE task.
Instead,  we extend the CRF training to leverage the confidence of the estimated posteriors.  
The new CRF training (called Uncertain Label Marginalizing (ULM)) treats low confidence tokens as missing labels and computes the sentence-level score by marginalizing over them. A similar idea has been previously used in treating partially labeled data~\cite{kim2015weakly}.

Specifically, given a sentence $\bm{x}$ we define a constrained \textit{lattice} $\mathcal{Y}(\bm{x})$, where at each position $t$ the allowed label types $\mathcal{Y}(x_t)$ are:
\begin{equation}
\mathcal{Y}(x_t) = 
\begin{cases}
\{y_t\}, \qquad \text{if} \quad p(y_t|\bm{x};\theta) > \eta\\
\text{All label types}, \quad \text{otherwise}
\end{cases}
\label{eq:ulm}
\end{equation}
where $\eta$ is the confidence threshold, $y_t$ is the prediction of posterior decoding and $p(y_t|\bm{x};\theta)$ is its CRF token marginal. The new neural network parameters $\theta$ are estimated by maximizing the log-likelihood of $p_\theta(\mathcal{Y}(\bm{x}^{k})|\bm{x}^{k})$ for every input sentence $\bm{x}^{k}$, where

\begin{equation*}
\label{eq:latscore1}
p_\theta(\mathcal{Y}(\bm{x}^{k})|\bm{x}^k) =  \frac{\sum_{\bm{y}^{k}\in \mathcal{Y}(\bm{x}^{k})}\exp(\phi(\bm{y}^k;\bm{x}^k,\theta))}{\sum_{\bm{y}'\in \bm{y}^m} \exp(\phi(\bm{y}';\bm{x},\theta))}
\end{equation*}
where $\bm{y}^{k}$ is an instance  sequence of lattice $\mathcal{Y}(\bm{x})$,  
and $k$ is the sentence index in the training set. 
Extreme cases are when all tokens are uncertain then the likelihood would be equal to 1, when all tokens of a sequence are confident, it would be equal to Eq. \ref{eq:score} where only one possible sequence, as in Fig.~\ref{fig:ULM}. 
\vspace{-.2cm}
\paragraph{Inductive and Transductive Learning} 
The semi-supervised training process is summarized as follow: It first computes marginals over the unlabeled data given  a set of CRF parameters. 
It then uses the marginals as a regularization term for label propagation. The smoothed posteriors from the graph are then interpolated with the CRF marginal in \textsc{GraphInterp} or used as an additional feature in \textsc{GraphFeat}. It then uses the estimated labels for  the unlabeled data combined with the labeled data to retrain the CRF using either the hard decision CRF training objective as Eq.~\ref{eq:score}  or the ULM data selection objective. 

In the inductive setting, we only use the unlabeled data from the development set for the semi-supervision, while in the transductive setting we need to use the unlabeled data of the test set to construct graph. In both cases, the parameters are tuned only on the dev set. 

\subsection{Experimental Results}
We  evaluate our \sys\ model in both supervised and semi-supervised settings on ScienceIE dataset. We also perform ablations and try different variants to best understand our model. Results are reported in \cite{luan2017scienceie}.
\begin{table}
  \centering
 {\footnotesize
  \begin{tabular}{lll}
    \toprule
  Span Level  & Identification & Classification  \\
  \midrule
    Gupta et.al.(unsupervised)& 9.8 & 6.4\\
    Tsai et.al. (unsupervised) & 11.9 & 8.0\\  \hdashline
    Best Non-Neural SemEval$^+$ & 38 & 51 \\
    Best Neural SemEval$^+$ & 44 & 56\\
    \sys(supervised) & 40.2 & 52.1\\
    \sys(semi)  & 45.3 & 56.9\\ \hdashline
    \sys(semi)$^*$ & \textbf{46.6} & 57.6\\
    \bottomrule
  \end{tabular}}
  \caption{ \small{Overall span-level F1 results for keyphrase identification (SemEval Subtask A) and classification (SemEval Subtask B). $^*$ indicates tranductive setting. $^+$ indicates not described in \cite{scienceIE} whether transductive or inductive. }}
  \label{tab:best_span}
\end{table}

 Table \ref{tab:best_span} reports the results of our neural sequence tagging model \sys\ in both  supervised  and semi-supervised learning (ULM and graph-based),  and compares them with the  baselines and the  state-of-the-art (best SemEval System~\cite{scienceIE}). We report results for both span identification (SemEval SubTask A) and span classification into \textsc{Task}, \textsc{Process} and \textsc{Material} (SemEval Subtask B).
  The results show that our neural sequence tagging models significantly outperforms the state of the art and both baselines.  It confirms that our neural tagging model outperforms other non-neural and neural models for the SemEval ScienceIE challenge.\footnote{Best SemEval Numbers from https://scienceie.github.io/}  
  It further shows that our system achieves significant boost from semi-supervised learning using unlabeled data.
 
 \begin{table}
  \centering
 {\footnotesize
   \begin{tabular}{ll|ll}
    \toprule
     Posterior & Training & Dev & Test\\
     \midrule
     - & -& 50.2 & 42.9 \\
     - &ULM & 51.3 & 44.4\\
     \textsc{GraphInterp}&- & 50.9 & 43.3\\
\textsc{GraphInterp} & ULM & \textbf{51.9} & \textbf{45.3} \\ \hline
               \textsc{GraphInterp}* &-& 50.7 & 44.0\\
\textsc{GraphInterp}* &ULM& 51.8 & 45.7\\
     \textsc{GraphFeat}* &- & 51.4 & 44.9 \\ 
\textsc{GraphFeat}* &ULM & \textbf{52.1} & \textbf{46.6} \\
    \bottomrule
  \end{tabular}}
  \caption{ \small{F1 scores of semi-supervised Learning approaches; * shows transductive models. }}
  \label{tab:ssl}
\end{table}

Table~\ref{tab:ssl} reports the results of the semi-supervised learning algorithms in different settings. In particular we ablate incorporating the graph-based methods of computing the posterior and CRF training (ULM vs. hard decision). The table shows incorporating graph-based methods for computing posterior and ULM for CRF training outperforms their counterparts. 

The transductive approaches consistently outperform inductive approaches on the test set. With around the same performance on dev set, \textsc{GraphInterp}* seems to generalize better on test set with 1.6\% relative improvement over \textsc{GraphInterp}. We observe higher improvement with \textsc{GraphFeat}* compared to \textsc{GraphInterp}. This is mainly because automatically learning the weight matrix $M$ between neural network scores and graph outputs adds more flexibility compared to tuning an interpolation weight $\alpha$.
 
  The performance is further improved by applying data selection through modifying the objective to ULM. The best inductive system is ULM+\textsc{GraphInterp} with 5.6\% relative improvement over pure self-training that makes hard decisions, and the best transductive system is ULM+\textsc{GraphFeat}* with 8.6\% relative improvement.
\section{Research Plan}
\subsection{Overview}
\label{sec:RP_overview}

We focus on building a structured knowledge graph that can show the relation between scientific terms and make recommendations based on user's query. For example, Generative Adversarial Networks (GAN) has been first introduced in machine learning communities in 2014, and reaches its peaks in 2015, yet is not applied  in the NLP field until 2017. NLP researchers may want to know the possible NLP applications of GAN, which could be inferred from relational learning of a scientific knowledge graph.  
Therefore, the goal of our research is to make scientific recommendations using more existing information and minimizing human annotation effort. We utilize semi-supervision, co-occurrence and signals from external resources to improve our system performance. The roadmap is as follows:

\begin{itemize}
\item \textbf{Entity Extraction and Linking}: Extract scientific terms from scientific articles and merge synonymous expressions of scientific terms.
\item \textbf{Knowledge Graph Construction}: Build a relational knowledge graph on a scientific domain without extensive human annotation effort.
\item \textbf{Scientific Recommendation}: Recommend the applications for a scientific method or recommend the methods that can solve a scientific application. This serves as an  evaluation of the graph.
\end{itemize}

\subsection{Data and Other Resources}
In addition to the ScienceIE data described in Sec.~\ref{sec:ScienceIE}, we choose AI as a broad area of focus  because our expertise can make it easier to analyze the results. We collect papers from several major AI communities to conduct the experiments.

\subsubsection{ACL Anthology Network (AAN) Dataset}
The ACL Anthology Network (AAN) Dataset  \cite{radev2009acl,bird2008acl} provides citation and collaboration networks of the articles included  in  the  ACL  Anthology, which consists of 23766 papers from 1965 to 2013.  
All papers in AAN are parsed by Parscit~\cite{councill2008parscit}, which parses scientific documents~\cite{luong2012logical} into the following logical structures:\textit{ abstract
, categories, general terms, keywords,  introduction, background, related work,  methodology, evaluation, discussion, conclusions, acknowledgments}. 

Based on AAN, Gupta and Manning \cite{gupta2011analyzing} first proposed a task that defines scientific terms into three aspects: \textit{domain}, \textit{technique} and \textit{focus} and hand-labeled 474 titles and abstracts in AAN with the three categories to measure the token-level precision and recall scores. Fig.~\ref{fig:sonal} shows an example of the annotation. We refer to the dataset as GM-ANN dataset.

\begin{figure}
\centering
\begin{footnotesize}
\begin{tabular}{p{14cm}}
\toprule
\textbf{[Discriminative Word Alignment]$\blue{\mathrm{_{\bm{Domain}}^{\bm{Focus}}}}$} via \textbf{[Alignment Matrix]$\blue{\mathrm{_{\bm{Technique}}^{\bm{Focus}}}}$} modelling.\\
In this paper a new \textbf{[discriminative Word Alignment]$\blue{\mathrm{_{\bm{Technique}}^{\bm{Focus}}}}$} method is presented.
This approach models directly the \textbf{[alignment matrix]$\blue{\mathrm{_{\bm{Technique}}}}$} by the \textbf{[conditional random field]$\blue{\mathrm{_{\bm{Technique}}}}$}  (\textbf{[CRF]$\blue{\mathrm{_{\bm{Technique}}}}$}) and so no restrictions to the alignments have to be made.
\\
\bottomrule
\end{tabular}
\end{footnotesize}
\footnotesize{\caption{\label{fig:sonal} Annotated \textit{Focus}, \textit{Domain}, \textit{Technique} examples from GM-ANN dataset \cite{gupta2011analyzing}.}}
\end{figure}

\subsubsection{AI2 dataset}

In order to have a broader coverage of research communities, we use the AI2 SemanticScholar Open Resource Corpus\footnote{http://labs.semanticscholar.org/corpus/} which has over 7 million published research papers in Computer Science and Neuroscience. 
We search through the corpus and collect conference proceedings from the following communities: (1) Machine Learning: NIPS and ICML (2) Speech: Interspeech and ICASSP, (3) Vision: CVPR, ICCV and ECCV (4) General AI: AAAI and IJCAI.
The statistics of the datasets are in Table \ref{tab:data}.

\begin{table}[t]
  \centering
 {\footnotesize
   \begin{tabular}{c|cc|c|cc|ccc|cc}
    \toprule
    
    Communities  & \multicolumn{2}{c|}{Machine Learning} & \multicolumn{1}{c|}{NLP} & \multicolumn{2}{c|}{Speech} & \multicolumn{3}{c|}{Vision} & \multicolumn{2}{c}{General AI}\\
        \midrule
     Venues &  NIPS & ICML & ANN & ICASSP & Interspeech &  CVPR & ICCV & ECCV & AAAI & IJCAI \\
    \midrule
     Paper \# & 6987 & 3078 & 23766 & 16576 & 8352 & 6615 & 2839 & 2227 & 4416 & 4799\\

    \bottomrule
  \end{tabular}}
  \caption{Statistics of venues we collect from both AI2 dataset and ACL anthology.}
  \label{tab:data}
\end{table}

The AI2 dataset has the information of  \textit{paper title}, \textit{paper abstract}, \textit{author}, \textit{inCitations}, \textit{outCitations},  \textit{venue} and \textit{year}. Even though the text of the full paper is not provided, the paper abstract and title are useful for our work.

\subsubsection{Wikipedia}
\begin{figure}[t]
\centering
\includegraphics[width=16cm]{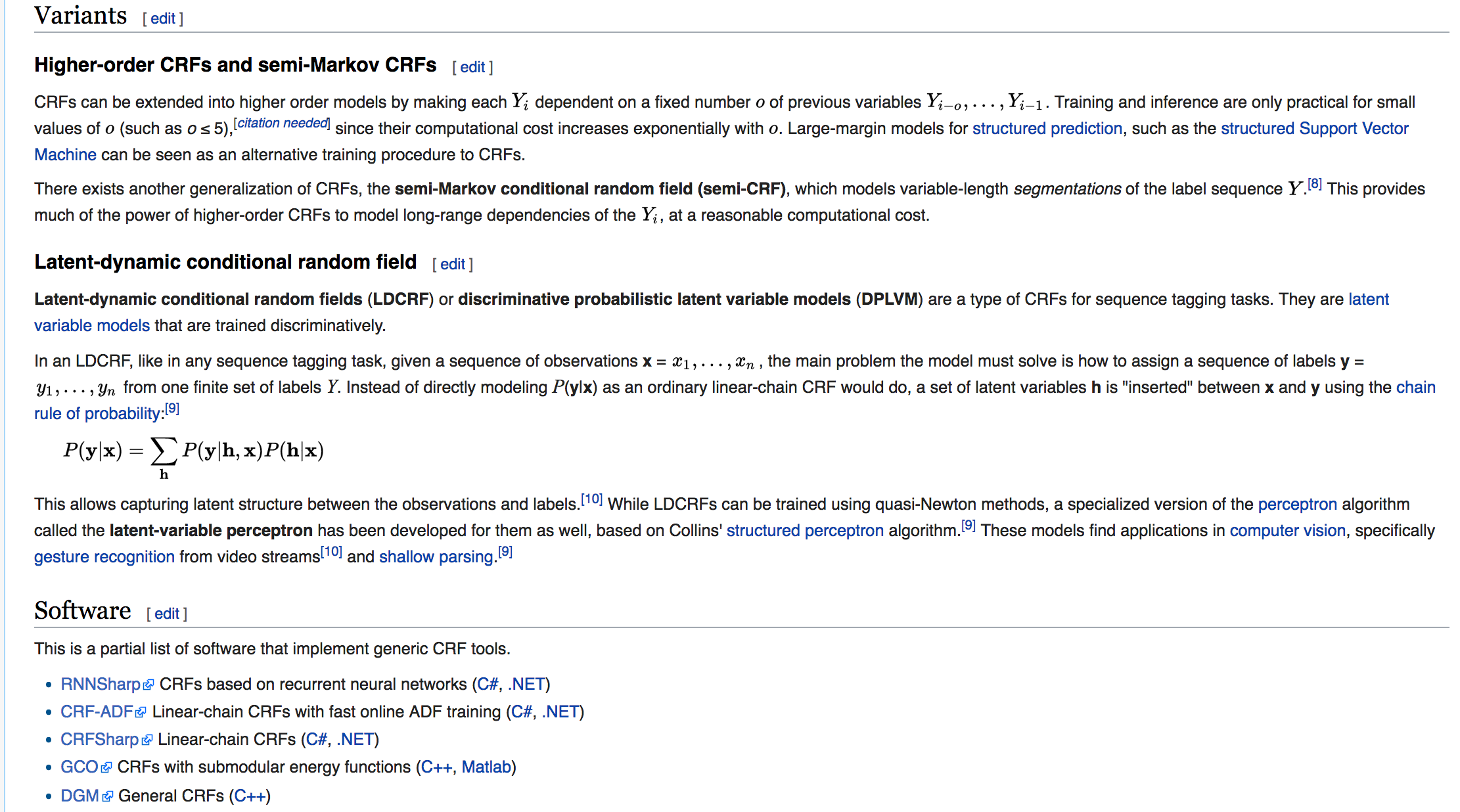}       
\caption{An example of Wikipedia page for term \textit{Conditional Random Field}}
\label{fig:wiki}
\end{figure}

Wikipedia includes a large collection of structural articles with the introduction of terms, where information about the term is categorized into sections with specific section titles. Moreover, hyperlinks of the phrases that are used to describe the term are provided, which means the keyphrase boundary can be identified by the hyperlink.  If the phrase existed in Wikipedia, a connection can be built through the hyperlink between the two terms.  Furthermore, the structural nature of Wikipedia provides a good resource of relations between scientific terms. Fig.~\ref{fig:wiki} provides an example of the Wikipedia page of term \textit{Conditional Random Field}, where there are two sections: Variants and Software. The terms under \textit{Variants} that are highlighted with bold fonts or hyperlinks are mostly the related methods such as \textit{structure prediction}, \textit{support vector machine} and \textit{latent variable models}. The terms  under \textit{Software} are all tools that can implement a CRF. In this way, we can extract relation triplets such as (\textit{support vector machine}, \textit{software}, \textit{Conditional Random Field}) and (\textit{RNNSharp}, \textit{Variant}, \textit{Conditional Random Field}).
The scientific terms can therefore connect to each other through the links and the contexts.

\subsection{Entity Extraction and Linking}
\label{sec:linking}
 All unannotated papers will be tagged by the extractor in Sec.~\ref{sec:pilot} trained using the ScienceIE and the GM-AAN dataset. The scientific terms extracted are classified into \textit{Task} and \textit{Method}.

\begin{figure}[t]
\centering
\includegraphics[width=15cm]{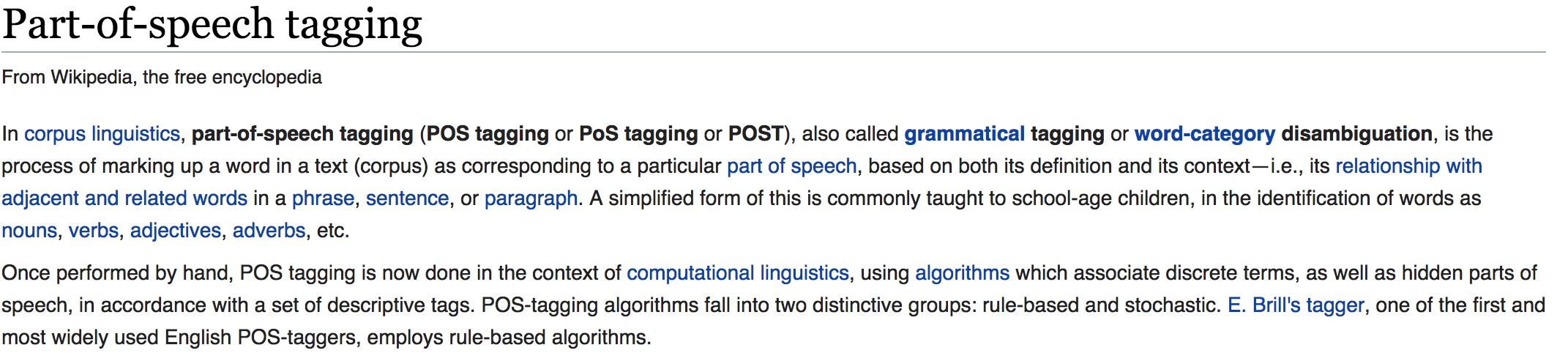}       
\caption{The description section of Wikipedia page for term \textit{Part-of-speech tagging}}
\label{fig:wiki_description}
\end{figure}
Once the terms are extracted, 
since there are different expressions for the same scientific term, we need to link them  to the same node in the knowledge graph. For example, \textit{part-of-speech tagging} can be called as \textit{POS tagging}, \textit{PoS tagging} or \textit{POST}. If part-of-speech tagging is used as a tool to solve other problems, it is also called as \textit{part-of-speech tagger}. 

Unlike general entity linking problems (such as linking human names) which have greater variation between entity mentions \cite{ling2015design,han2011generative},  spelling variants and acronyms play the central role in disambiguating the scientific terms. We will apply knowledge-driven feature extraction with simple classifier and rules and use the ScienceIE synonym relations for evaluation.

\subsection{Knowledge Graph Construction}
\label{sec:relation}
In order to minimize the effort for human annotation, we aim at exploring relational signals embedded in the text such as co-occurrence and dependency relations,  as well as relational signals from external resources such as Wikipedia. We first construct a knowledge graph using a set of fixed relation types obtained from term co-occurrence, then add auxiliary relations to improve the graph.
We also propose possible ways of modeling open relations through latent representations. 

\subsubsection{Relational Learning through Term Co-occurrence}
Two scientific terms can be assumed to be related if they co-occur in a context window (sentence, paragraph, section or document). For example, as in Figure \ref{fig:sonal}, the term  \textit{Alignment Matrix} is an approach to solve the problem of  \textit{Discriminative Word Alignment}. The two terms appears in the same sentence multiple times which is a strong indicator for them to be related to each other. 

A traditional way of getting the relation category is to train a relation extractor which can classify relation of any two terms into predefined  type of relations through context. Since we do not have annotation for the relation extractor, we use the entity types together with occurrence to infer relations. 
The entity extractor  outputs the keyword boundary as well as their categories: each keyword is marked as \textit{Task} and \textit{Method}. If a term is labeled as \textit{Task}, and another term in the same context window labeled as \textit{Method}, we can assume that the two terms are \textit{Task-Method} relation (the second term is a method used to solve the first term), such as \textit{Discriminative Word Alignment} and \textit{Alignment Matrix} in Fig.~\ref{fig:sonal}. 

Based on the keyword categories, we define three broad relation types: \textit{Task-Task} which indicates the two scientific terms are related tasks, \textit{Task-Method} which indicates that one term is a task that can be solve by the other,  \textit{Method-Method} which indicates the two terms are related methods. We can then extract relation triplets from term co-occurrence, such as \{\textit{Discriminative Word Alignment}, \textit{Task-Method}, \textit{Alignment Matrix}\} in the example above. Since the output of term extracter is noisy, we can reduce noise by filtering out the low confident terms.

 We follow previous work on knowledge graph inference in \cite{yang2014embedding} as described in Sec.~\ref{sec:singleKB}, and use bilinear transformation (Eq.~\ref{eq:bilinear}) as our scoring function. The model will be trained by minimizing the ranking loss (Eq.~\ref{eq:objective}) of all triplets extracted from co-occurrence. Note that even though the triplets extracted from co-occurrence are noisy, the objective can naturally put more weight on the more frequent triplets and learn parameters accordingly.

\subsubsection{Leveraging Auxiliary Relations}
In order to reduce the noise from automatic annotation by co-occurrence, we will use  auxiliary relations from dependency paths between the scientific terms as in many previous studies\cite{riedel2013relation,toutanova2016compositional}. In addition, we will also introduce structural external resources such as Wikipedia to augment the graph. 

\paragraph{Dependency Paths}
Syntactic structure provides an important clue for modeling relations, and many state-of-the-art systems use dependency relation as an auxiliary information for KB completion \cite{riedel2013relation,toutanova2016compositional}. 
In scientific literature, related terms are usually embedded across sentences. Take the second and the third sentence in Fig.~\ref{fig:sonal} as an example, \textit{discriminative Word Alignment} is connected to \textit{alignment matrix} through the coreference  indicator \textit{This approach}. Therefore it is very important to consider contexts beyond sentence boundaries when using dependency paths to augment the graph. Following \cite{peng2017cross}, we will incorporate both intra-sentential and inter-sentential dependencies, such as sequential, syntactic,  coreference and  discourse  relations as auxiliary relations using the tool of \cite{peng2017cross}.

\begin{figure}[t]
\centering
\includegraphics[width=12.5cm]{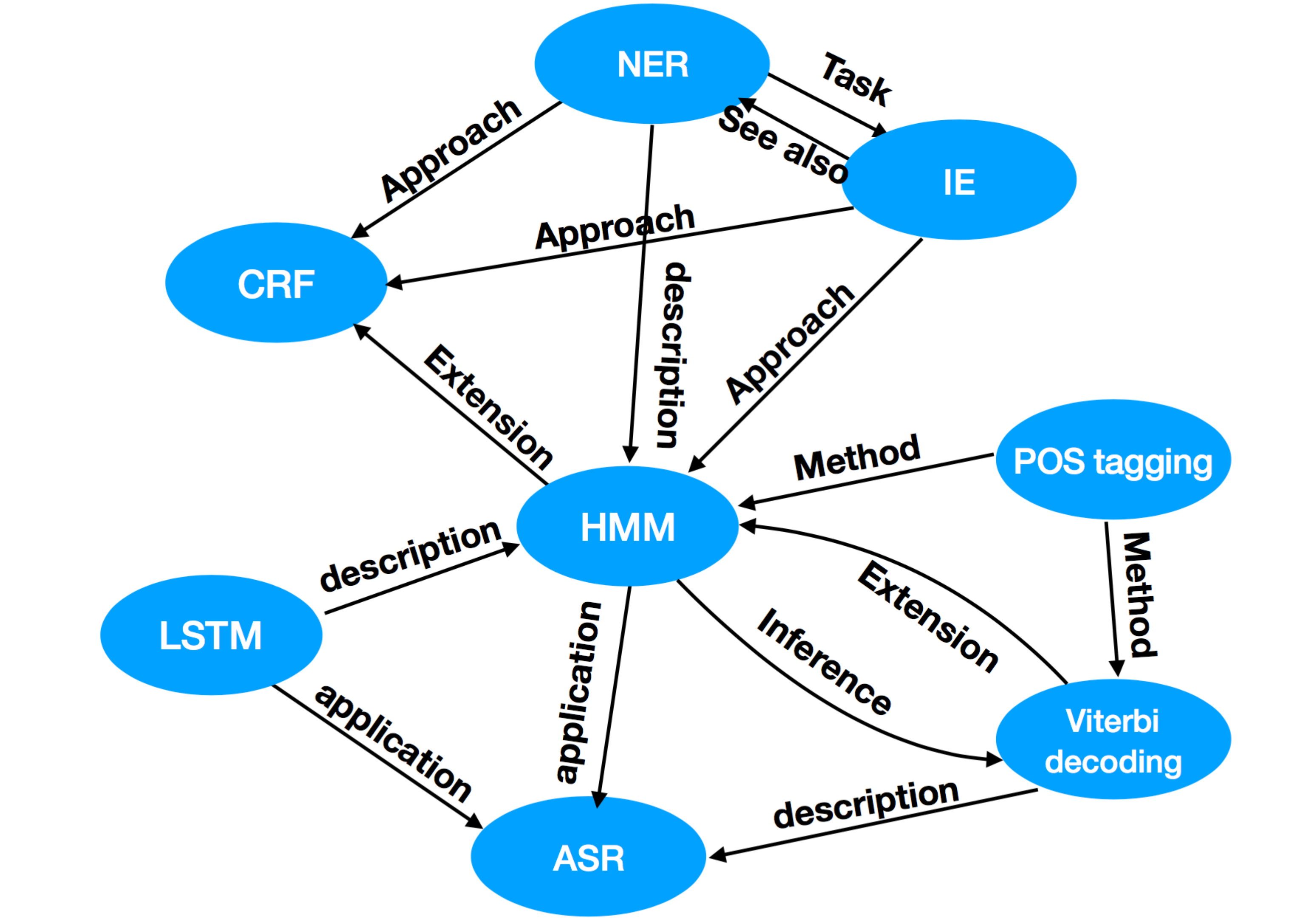}       
\caption{Wikipedia graph: The arrow starts from the Wikipedia page where the relation triplet is extracted. The arrow points to  the scientific term that appears in the page. The section that the term appears in is shown by the arch caption. }
\label{fig:wiki_graph}
\end{figure}

\paragraph{Wikipedia}
The internal links between Wikipedia page formulate a graph connecting scientific terms.   Moreover, the section  where the link appears in is a strong indicator for the relation between scientific terms. For example, the term \textit{latent variable models} appears under \textit{Variants} section of the Wikipedia page \textit{Conditional Random Field} in Figure \ref{fig:wiki}, \textit{latent variable models} and \textit{Conditional Random Field} are likely to be similar methods that have many common applications.   
We will use a set of heuristics to define a set of categories  on the major section names of Wikipedia, such as  \textit{Description}, \textit{Extension}, \textit{Example}, \textit{Application}, \textit{Models and Methods}, \textit{Tools},  \textit{Approach}, \textit{See also} and so on. If a term appears on another term's Wikipedia page under the section of \textit{Application}, the two terms are likely to be \textit{Method-Task} relation. The section names can help distinguish the relation, either explicitly (such as \textit{Application} and \textit{Models and Methods}), or implicitly (such as \textit{See also} and \textit{Extension}).  An example of a Wikipedia-derived graph is shown in Fig.~\ref{fig:wiki_graph}.

\begin{figure}[t]
\centering
\includegraphics[width=11cm]{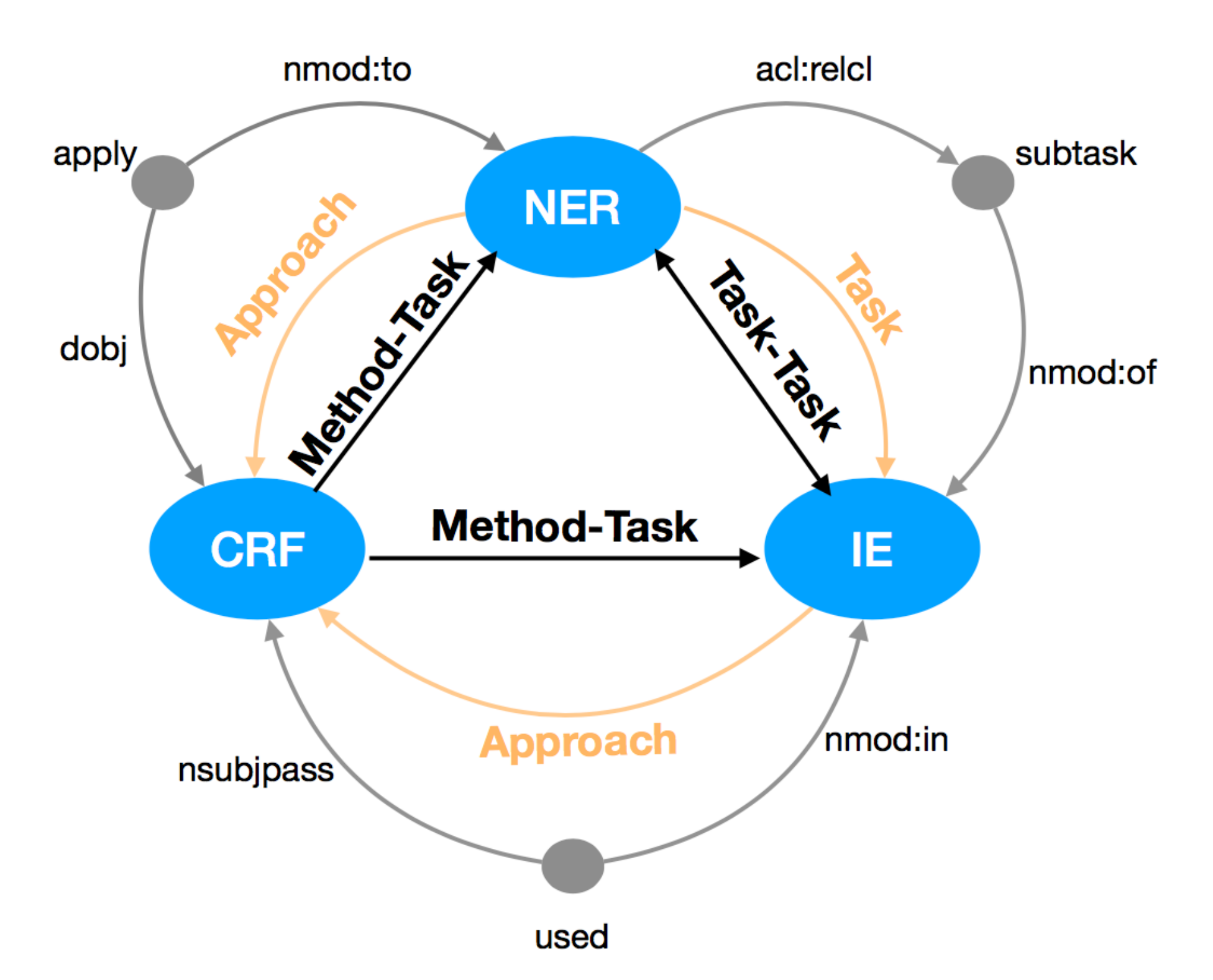}       
\caption{Graph with auxiliary relations from different resources: Yellow arches are relations from Wikipedia sections; Grey arches are dependency relations extracted from text where the two terms co-occur. Black arches are scientific aspect relations (the main relation  we focus on).}
\label{fig:external_graph}
\end{figure}

Figure \ref{fig:external_graph} is an example of a graph with auxiliary relations from different resources. 
As shown in the figure, some auxiliary relations connects two scientific terms through multiple nodes (for example the dependency relations). Many previous studies showed that multiple-step relation paths also contain rich inference patterns between entities, and yield significant gains in embedding models for knowledge base completion tasks \cite{lin2015modeling,guu2015traversing,toutanova2016compositional}.
Therefore, it is necessary to model multiple-step relation paths through compositional methods.

 In order to model multiple-step relation paths, we follow the the approach of \cite{toutanova2016compositional} as described in Sec.~\ref{sec:multiKB}. 
 We redefine the relation matrix $Q_r$ in Eq.~\ref{eq:bilinear} as  $Q^{(k)}_r$, which means the $r$-th relation type on the $k$-th auxiliary resource of  knowledge.
For each type of resource $k$, we extract relation triplets between two scientific terms $\{e_x,\pi^{(k)}, e_y\}$, where $\pi^{(k)}$ is the path between the two terms which may consists of multiple steps. We take into account embeddings of relation paths between the two terms, and use vector space compositions to combine the embeddings of individual relation links $r_i$ 
into an embedding of the path $\pi^{(k)}$. The intermediate nodes $e_i$ are neglected. 

We calculate the scoring function in Eq.~\ref{eq:bilinear_multi} for path $\pi^{(k)}$ from each auxiliary resource $k$:

\begin{equation}
\phi(\pi^{(k)}) = Q_{r_1}^{(k)} \dots Q_{r_n}^{(k)}
\end{equation}

The weighted path representation will be 
\begin{equation}
\label{eq:path}
F(e_x,e_y, r)=\sum_k\sum_{\pi^{(k)}} w_{|\pi|}^{(k)}P(e_y|e_x,\pi^{(k)})\phi(\pi^{(k)}),
\end{equation}
 where $w_{|\pi|}^{(k)}$ is a shared parameter for paths of each length for each resource type $k$, so that the model may learn to trust the contribution of different path lengths for different resources. $P(e_y|e_x,\pi^{(k)})$ is the random walk probability for each path.
We use the same scoring function as Eq.~\ref{eq:score_multi} and objective function as Eq.~\ref{eq:objective}.
We hope  through multi-path modeling of multiple resources, the auxiliary relations would improve the performance of scientific recommendation.

\subsubsection{Extension to open relation learning}
In the proposed method above, we only define three broad target relations for recommendations, which is not enough to satisfy all users' query. Some more fine-grained relations that are under the three broad relations may be more useful. For example, two related methods can both be applied to solve one task, and their performance is usually compared, such as \textit{Logistic Regression} and \textit{SVM}. On the other hand, one method may be combined with another method to solve a task, such as \textit{HMM} and \textit{Viterbi Decoding}. Similarly, two tasks can be hypernyms such as \textit{IE} and \textit{NER}, some tasks can be similar problems that can involve the same methods, such as \textit{POS tagging} and \textit{NER}. 

Even though we do not have any annotation for those relations, we can learn latent representations through clustering the  representations, e.g. the weighted path representation in Eq. ~\ref{eq:path}. It can also be obtained from matrix or tensor based low rank factorization. For analysis, we can choose some examples around the cluster centroid and use a small amount of human effort to label the relation each cluster represents.

\subsection{Scientific Recommendation}

\begin{figure}[t]
\centering
\includegraphics[width=12cm]{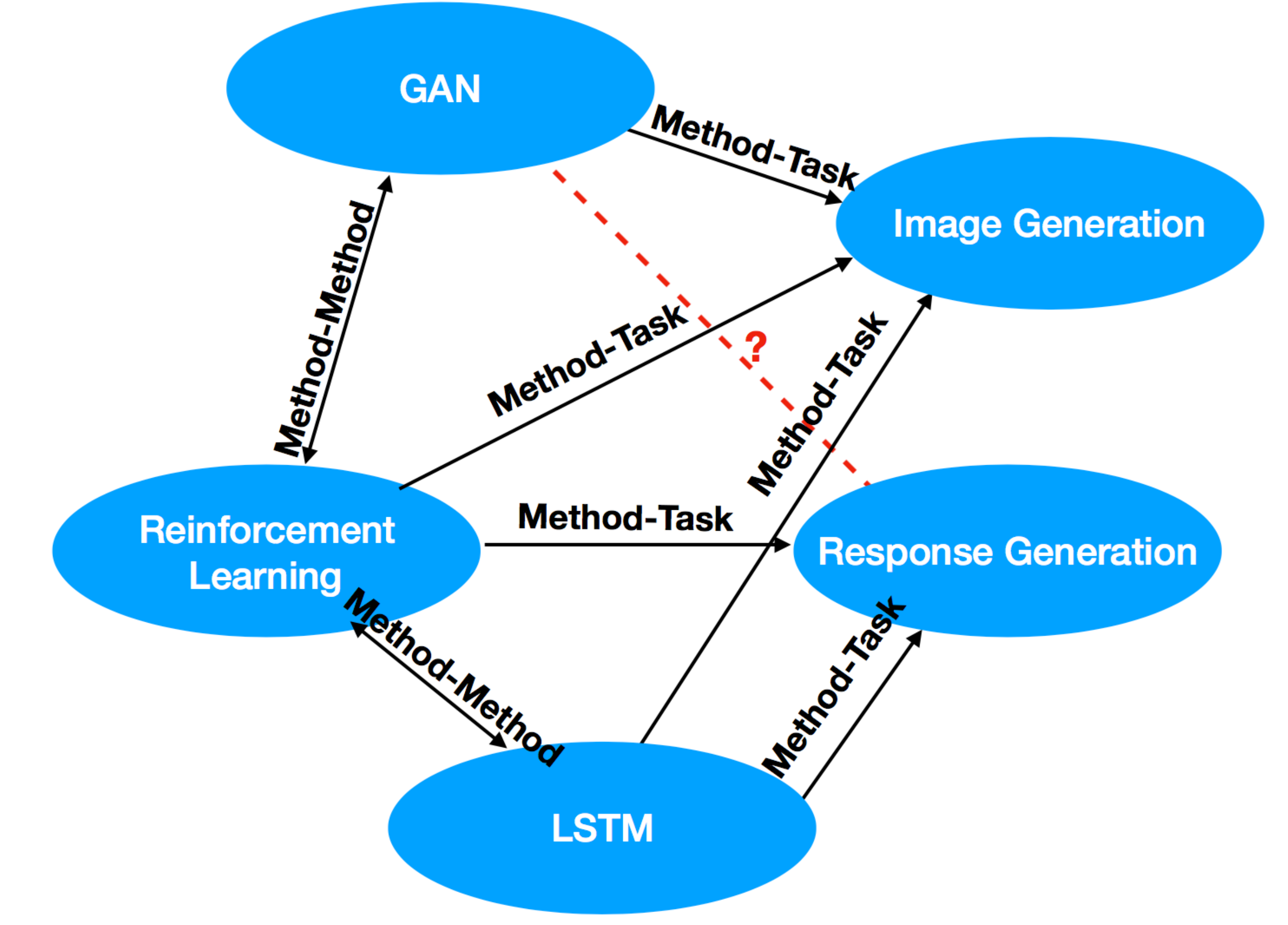}       
\caption{Link Prediction for scientific terms. The relation between \textit{GAN} and \textit{Response Generation} can be inferred from the graph with a ranking score. }
\label{fig:term_relation}
\end{figure}

Our system will be applied to recommend related tasks or methods to a new scientific term.
For example, since \textit{GAN}  is a new technique that has not been widely used in NLP field until 2017,  we can set our task as predicting the possible applications of \textit{GAN} in NLP field as Fig.~\ref{fig:term_relation}. 
The problem is similar to link prediction \cite{toutanova2016compositional} and knowledge base completion, and \cite{trouillon2017knowledge} can be formulated as a ranking problem where the  score between \textit{GAN} and all other scientific terms with the relation of \textit{Method-Task} are calculated and ranked in descending order.

For automatic evaluation, we will hold out all publications about a set of new methods such as \textit{GAN}  in ACL anthology as test set, all publication in ACL anthology 2016 as dev set and  use all AI publications (NIPs, CVPR, ACL, etc.) prior to 2016 as a training set to build the knowledge graph.
The predicted \textit{GAN} applications are compared with the \textit{GAN} papers in 2017, where ground truth ranking is calculated by the number of papers the term appears.  The system performance  can thus be evaluated by Mean Reciprocal Rank (MRR) or \textit{Hits  at k}.

The automatic evaluation has the limitation that there may be some applications that are indeed  good applications for \textit{GAN}, but have not been published yet. Therefore, in addition to automatic evaluation, we introduce human evaluation. Inspired by TREC competitions \cite{schutze2008introduction} and \cite{riedel2013relation}, we treat the set of
new methods (or tasks) as query and receive the top 30 candidate scientific terms from each system
and manually judge their relevance. This
gives a set of relevant results that we can use to calculate
recall and precision measures. 
\section{Conclusion}
In summary, we propose a  framework that can make scientific recommendations using minimum effort of human annotation. The anticipated contributions of this thesis work can be summarized as follows. First, we  obtained state-of-the-art performance on scientific term extraction by leveraging a large amount of unannotated papers through applying multiple semi-supervised approaches. Secondly, we construct a scientific knowledge graph by combining relational signals embedded in the text such as co-occurrence with auxiliary relations from dependency relations or Wikipedia. We also explore possible ways of representing open relations through clustering latent relational representations. Finally, we propose downstream scientific recommendation tasks that will be of interest to many researchers. 
Evaluation schemes which can minimize annotation efforts are also proposed. Our approaches  to minimize annotation efforts can be generalized to other domains and may provide new solutions to problems in machine reading.
\bibliographystyle{IEEEbib}
\bibliography{references}

\begin{thebibliography}{10}

\bibitem{luan2018uwnlp}
Yi~Luan, Mari Ostendorf, and Hannaneh Hajishirzi,
\newblock ``The {UWNLP} system at semeval-2018 task 7: Neural relation
  extraction model with selectively incorporated concept embeddings,''
\newblock in {\em Proc.\ Int.\ Workshop on Semantic Evaluation (SemEval)},
  2018, pp. 788--792.

\bibitem{luan2018multi}
Yi~Luan, Luheng He, Mari Ostendorf, and Hannaneh Hajishirzi,
\newblock ``Multi-task identification of entities, relations, and coreference
  for scientific knowledge graph construction,''
\newblock in {\em Proc.\ Conf. Empirical Methods Natural Language Process.
  (EMNLP)}, 2018.

\bibitem{hixon2015learning}
Ben Hixon, Peter Clark, and Hannaneh Hajishirzi,
\newblock ``Learning knowledge graphs for question answering through
  conversational dialog.,''
\newblock in {\em HLT-NAACL}, 2015, pp. 851--861.

\bibitem{waltz1978english}
David~L Waltz,
\newblock ``An english language question answering system for a large
  relational database,''
\newblock {\em Communications of the ACM}, vol. 21, no. 7, pp. 526--539, 1978.

\bibitem{belleau2008bio2rdf}
Fran{\c{c}}ois Belleau, Marc-Alexandre Nolin, Nicole Tourigny, Philippe
  Rigault, and Jean Morissette,
\newblock ``Bio2rdf: towards a mashup to build bioinformatics knowledge
  systems,''
\newblock {\em Journal of biomedical informatics}, vol. 41, no. 5, pp.
  706--716, 2008.

\bibitem{ruttenberg2009life}
Alan Ruttenberg, Jonathan~A Rees, Matthias Samwald, and M~Scott Marshall,
\newblock ``Life sciences on the semantic web: the neurocommons and beyond,''
\newblock {\em Briefings in bioinformatics}, vol. 10, no. 2, pp. 193--204,
  2009.

\bibitem{toutanova2016compositional}
Kristina Toutanova, Xi~Victoria Lin, Wen-tau Yih, Hoifung Poon, and Chris
  Quirk,
\newblock ``Compositional learning of embeddings for relation paths in
  knowledge bases and text,''
\newblock in {\em Proceedings of the 54th Annual Meeting of the Association for
  Computational Linguistics}, 2016, vol.~1, pp. 1434--1444.

\bibitem{quirk2016distant}
Chris Quirk and Hoifung Poon,
\newblock ``Distant supervision for relation extraction beyond the sentence
  boundary,''
\newblock {\em arXiv preprint arXiv:1609.04873}, 2016.

\bibitem{etzioni2008open}
Oren Etzioni, Michele Banko, Stephen Soderland, and Daniel~S Weld,
\newblock ``Open information extraction from the web,''
\newblock {\em Communications of the ACM}, vol. 51, no. 12, pp. 68--74, 2008.

\bibitem{collobert2011natural}
Ronan Collobert, Jason Weston, L{\'e}on Bottou, Michael Karlen, Koray
  Kavukcuoglu, and Pavel Kuksa,
\newblock ``Natural language processing (almost) from scratch,''
\newblock {\em Journal of Machine Learning Research}, vol. 12, no. Aug, pp.
  2493--2537, 2011.

\bibitem{saha2008gazetteer}
Sujan~Kumar Saha, Sudeshna Sarkar, and Pabitra Mitra,
\newblock ``Gazetteer preparation for named entity recognition in indian
  languages.,''
\newblock in {\em IJCNLp}, 2008, pp. 9--16.

\bibitem{settles2004biomedical}
Burr Settles,
\newblock ``Biomedical named entity recognition using conditional random fields
  and rich feature sets,''
\newblock in {\em Proceedings of the International Joint Workshop on Natural
  Language Processing in Biomedicine and its Applications}. Association for
  Computational Linguistics, 2004, pp. 104--107.

\bibitem{luan2016lstm}
Yi~Luan, Yangfeng Ji, and Mari Ostendorf,
\newblock ``{LSTM} based conversation models,''
\newblock in {\em arXiv preprint arXiv:1603.09457}, 2016.

\bibitem{luan2015efficient}
Yi~Luan, Shinji Watanabe, and Bret Harsham,
\newblock ``Efficient learning for spoken language understanding tasks with
  word embedding based pre-training,''
\newblock in {\em Proc.\ Conf. Int. Speech Communication Assoc. (INTERSPEECH)}.
  Citeseer, 2015, pp. 1398--1402.

\bibitem{lample2016neural}
Guillaume Lample, Miguel Ballesteros, Sandeep Subramanian, Kazuya Kawakami, and
  Chris Dyer,
\newblock ``Neural architectures for named entity recognition,''
\newblock {\em arXiv preprint arXiv:1603.01360}, 2016.

\bibitem{peng2015named}
Nanyun Peng and Mark Dredze,
\newblock ``Named entity recognition for chinese social media with jointly
  trained embeddings.,''
\newblock in {\em EMNLP}, 2015, pp. 548--554.

\bibitem{lafferty2001conditional}
John Lafferty, Andrew McCallum, Fernando Pereira, et~al.,
\newblock ``Conditional random fields: Probabilistic models for segmenting and
  labeling sequence data,''
\newblock in {\em Proceedings of the eighteenth international conference on
  machine learning, ICML}, 2001, vol.~1, pp. 282--289.

\bibitem{huang2015bidirectional}
Zhiheng Huang, Wei Xu, and Kai Yu,
\newblock ``Bidirectional lstm-crf models for sequence tagging,''
\newblock {\em arXiv preprint arXiv:1508.01991}, 2015.

\bibitem{hochreiter1997long}
Sepp Hochreiter and J{\"u}rgen Schmidhuber,
\newblock ``Long short-term memory,''
\newblock {\em Neural computation}, vol. 9, no. 8, pp. 1735--1780, 1997.

\bibitem{chiu2015named}
Jason~PC Chiu and Eric Nichols,
\newblock ``Named entity recognition with bidirectional lstm-cnns,''
\newblock {\em arXiv preprint arXiv:1511.08308}, 2015.

\bibitem{ballesteros2015improved}
Miguel Ballesteros, Chris Dyer, and Noah~A Smith,
\newblock ``Improved transition-based parsing by modeling characters instead of
  words with lstms,''
\newblock {\em arXiv preprint arXiv:1508.00657}, 2015.

\bibitem{ma2016end}
Xuezhe Ma and Eduard Hovy,
\newblock ``End-to-end sequence labeling via bi-directional lstm-cnns-crf,''
\newblock {\em arXiv preprint arXiv:1603.01354}, 2016.

\bibitem{etzioni2005unsupervised}
Oren Etzioni, Michael Cafarella, Doug Downey, Ana-Maria Popescu, Tal Shaked,
  Stephen Soderland, Daniel~S Weld, and Alexander Yates,
\newblock ``Unsupervised named-entity extraction from the web: An experimental
  study,''
\newblock {\em Artificial intelligence}, vol. 165, no. 1, pp. 91--134, 2005.

\bibitem{banko2007open}
Michele Banko, Michael~J Cafarella, Stephen Soderland, Matthew Broadhead, and
  Oren Etzioni,
\newblock ``Open information extraction from the web.,''
\newblock in {\em IJCAI}, 2007, vol.~7, pp. 2670--2676.

\bibitem{yang2014embedding}
Bishan Yang, Wen-tau Yih, Xiaodong He, Jianfeng Gao, and Li~Deng,
\newblock ``Embedding entities and relations for learning and inference in
  knowledge bases,''
\newblock {\em arXiv preprint arXiv:1412.6575}, 2014.

\bibitem{nickel2011three}
Maximilian Nickel, Volker Tresp, and Hans-Peter Kriegel,
\newblock ``A three-way model for collective learning on multi-relational
  data,''
\newblock in {\em Proceedings of the 28th international conference on machine
  learning (ICML-11)}, 2011, pp. 809--816.

\bibitem{riedel2013relation}
Sebastian Riedel, Limin Yao, Andrew McCallum, and Benjamin~M Marlin,
\newblock ``Relation extraction with matrix factorization and universal
  schemas,''
\newblock 2010.

\bibitem{guu2015traversing}
Kelvin Guu, John Miller, and Percy Liang,
\newblock ``Traversing knowledge graphs in vector space,''
\newblock {\em arXiv preprint arXiv:1506.01094}, 2015.

\bibitem{lin2015modeling}
Yankai Lin, Zhiyuan Liu, Huanbo Luan, Maosong Sun, Siwei Rao, and Song Liu,
\newblock ``Modeling relation paths for representation learning of knowledge
  bases,''
\newblock {\em arXiv preprint arXiv:1506.00379}, 2015.

\bibitem{schlichtkrull2017modeling}
Michael Schlichtkrull, Thomas~N Kipf, Peter Bloem, Rianne van~den Berg, Ivan
  Titov, and Max Welling,
\newblock ``Modeling relational data with graph convolutional networks,''
\newblock {\em arXiv preprint arXiv:1703.06103}, 2017.

\bibitem{athar2012detection}
Awais Athar and Simone Teufel,
\newblock ``Detection of implicit citations for sentiment detection,''
\newblock in {\em Proceedings of the Workshop on Detecting Structure in
  Scholarly Discourse}. Association for Computational Linguistics, 2012, pp.
  18--26.

\bibitem{athar2012context}
Awais Athar and Simone Teufel,
\newblock ``Context-enhanced citation sentiment detection,''
\newblock in {\em Proceedings of the 2012 conference of the North American
  chapter of the Association for Computational Linguistics: Human language
  technologies}. Association for Computational Linguistics, 2012, pp. 597--601.

\bibitem{do2013extracting}
Huy Hoang~Nhat Do, Muthu~Kumar Chandrasekaran, Philip~S Cho, and Min~Yen Kan,
\newblock ``Extracting and matching authors and affiliations in scholarly
  documents,''
\newblock in {\em Proceedings of the 13th ACM/IEEE-CS joint conference on
  Digital libraries}. ACM, 2013, pp. 219--228.

\bibitem{jaidka2014computational}
Kokil Jaidka, Muthu~Kumar Chandrasekaran, Beatriz~Fisas Elizalde, Rahul Jha,
  Christopher Jones, Min-Yen Kan, Ankur Khanna, Diego Molla-Aliod, Dragomir~R
  Radev, Francesco Ronzano, et~al.,
\newblock ``The computational linguistics summarization pilot task,''
\newblock {\em Proceedings of TAC}, 2014.

\bibitem{kas2011structures}
Miray Kas,
\newblock ``Structures and statistics of citation networks,''
\newblock Tech. {R}ep., DTIC Document, 2011.

\bibitem{sim2012discovering}
Yanchuan Sim, Noah~A Smith, and David~A Smith,
\newblock ``Discovering factions in the computational linguistics community,''
\newblock in {\em Proceedings of the ACL-2012 Special Workshop on Rediscovering
  50 Years of Discoveries}. Association for Computational Linguistics, 2012,
  pp. 22--32.

\bibitem{vogel2012he}
Adam Vogel and Dan Jurafsky,
\newblock ``He said, she said: Gender in the acl anthology,''
\newblock in {\em Proceedings of the ACL-2012 Special Workshop on Rediscovering
  50 Years of Discoveries}. Association for Computational Linguistics, 2012,
  pp. 33--41.

\bibitem{prabhakaran2016predicting}
Vinodkumar Prabhakaran, William~L Hamilton, Daniel~A McFarland, and Dan
  Jurafsky,
\newblock ``Predicting the rise and fall of scientific topics from trends in
  their rhetorical framing.,''
\newblock in {\em ACL (1)}, 2016.

\bibitem{anderson2012towards}
Ashton Anderson, Dan McFarland, and Dan Jurafsky,
\newblock ``Towards a computational history of the acl: 1980-2008,''
\newblock in {\em Proceedings of the ACL-2012 Special Workshop on Rediscovering
  50 Years of Discoveries}. Association for Computational Linguistics, 2012,
  pp. 13--21.

\bibitem{abu2011coherent}
Amjad Abu-Jbara and Dragomir Radev,
\newblock ``Coherent citation-based summarization of scientific papers,''
\newblock in {\em Proceedings of the 49th Annual Meeting of the Association for
  Computational Linguistics: Human Language Technologies-Volume 1}. Association
  for Computational Linguistics, 2011, pp. 500--509.

\bibitem{gupta2011analyzing}
Sonal Gupta and Christopher~D Manning,
\newblock ``Analyzing the dynamics of research by extracting key aspects of
  scientific papers.,''
\newblock in {\em IJCNLP}, 2011, pp. 1--9.

\bibitem{tsai2013concept}
Chen-Tse Tsai, Gourab Kundu, and Dan Roth,
\newblock ``Concept-based analysis of scientific literature,''
\newblock in {\em Proceedings of the 22nd ACM international conference on
  Conference on information \& knowledge management}. ACM, 2013, pp.
  1733--1738.

\bibitem{collins1999unsupervised}
Michael Collins and Yoram Singer,
\newblock ``Unsupervised models for named entity classification,''
\newblock in {\em Proceedings of the joint SIGDAT conference on empirical
  methods in natural language processing and very large corpora}. Citeseer,
  1999, pp. 100--110.

\bibitem{dienstmann2015database}
Rodrigo Dienstmann, In~Sock Jang, Brian Bot, Stephen Friend, and Justin
  Guinney,
\newblock ``Database of genomic biomarkers for cancer drugs and clinical
  targetability in solid tumors,''
\newblock {\em Cancer discovery}, vol. 5, no. 2, pp. 118--123, 2015.

\bibitem{peng2017cross}
Nanyun Peng, Hoifung Poon, Chris Quirk, Kristina Toutanova, and Wen-tau Yih,
\newblock ``Cross-sentence n-ary relation extraction with graph lstms,''
\newblock {\em Transactions of the Association for Computational Linguistics},
  vol. 5, pp. 101--115, 2017.

\bibitem{gerber2010beyond}
Matthew Gerber and Joyce~Y Chai,
\newblock ``Beyond nombank: A study of implicit arguments for nominal
  predicates,''
\newblock in {\em Proceedings of the 48th Annual Meeting of the Association for
  Computational Linguistics}. Association for Computational Linguistics, 2010,
  pp. 1583--1592.

\bibitem{yoshikawa2011coreference}
Katsumasa Yoshikawa, Sebastian Riedel, Tsutomu Hirao, Masayuki Asahara, and
  Yuji Matsumoto,
\newblock ``Coreference based event-argument relation extraction on biomedical
  text,''
\newblock {\em Journal of Biomedical Semantics}, vol. 2, no. 5, pp. S6, 2011.

\bibitem{teufel2002summarizing}
Simone Teufel and Marc Moens,
\newblock ``Summarizing scientific articles: experiments with relevance and
  rhetorical status,''
\newblock {\em Computational linguistics}, vol. 28, no. 4, pp. 409--445, 2002.

\bibitem{de2012verb}
Anita de~Waard and Henk~Pander Maat,
\newblock ``Verb form indicates discourse segment type in biological research
  papers: Experimental evidence,''
\newblock {\em Journal of English for academic purposes}, vol. 11, no. 4, pp.
  357--366, 2012.

\bibitem{dai2015semi}
Andrew~M Dai and Quoc~V Le,
\newblock ``Semi-supervised sequence learning,''
\newblock in {\em Advances in Neural Information Processing Systems}, 2015, pp.
  3079--3087.

\bibitem{luan2014semi}
Yi~Luan, Daisuke Saito, Yosuke Kashiwagi, Nobuaki Minematsu, and Keikichi
  Hirose,
\newblock ``Semi-supervised noise dictionary adaptation for exemplar-based
  noise robust speech recognition,''
\newblock in {\em Proc. Int. Conf. Acoustic, Speech, and Signal Process.
  (ICASSP)}. IEEE, 2014, pp. 1745--1748.

\bibitem{collobert2008unified}
Ronan Collobert and Jason Weston,
\newblock ``A unified architecture for natural language processing: Deep neural
  networks with multitask learning,''
\newblock in {\em Proceedings of the 25th international conference on Machine
  learning}. ACM, 2008, pp. 160--167.

\bibitem{luan2017multi}
Yi~Luan, Chris Brockett, Bill Dolan, Jianfeng Gao, and Michel Galley,
\newblock ``Multi-task learning for speaker-role adaptation in neural
  conversation models.,''
\newblock in {\em Proc.\ Int. Joint Conf. on Natural Language Processing
  (IJCNLP)}, 2017.

\bibitem{mikolov2013efficient}
Tomas Mikolov, Kai Chen, Greg Corrado, and Jeffrey Dean,
\newblock ``Efficient estimation of word representations in vector space,''
\newblock {\em arXiv preprint arXiv:1301.3781}, 2013.

\bibitem{pennington2014glove}
Jeffrey Pennington, Richard Socher, and Christopher~D Manning,
\newblock ``Glove: Global vectors for word representation.,''
\newblock in {\em EMNLP}, 2014, vol.~14, pp. 1532--1543.

\bibitem{levy2014dependency}
Omer Levy and Yoav Goldberg,
\newblock ``Dependency-based word embeddings.,''
\newblock in {\em ACL (2)}. Citeseer, 2014, pp. 302--308.

\bibitem{luan2016multiplicative}
Yi~Luan, Yangfeng Ji, Hannaneh Hajishirzi, and Boyang Li,
\newblock ``Multiplicative representations for unsupervised semantic role
  induction,''
\newblock in {\em Proc.\ Annu. Meeting Assoc. for Computational Linguistics
  (ACL)}, 2016, p. 118.

\bibitem{nigam2000text}
Kamal Nigam, Andrew~Kachites McCallum, Sebastian Thrun, and Tom Mitchell,
\newblock ``Text classification from labeled and unlabeled documents using
  em,''
\newblock {\em Machine learning}, vol. 39, no. 2, pp. 103--134, 2000.

\bibitem{baluja1998probabilistic}
Shumeet Baluja,
\newblock ``Probabilistic modeling for face orientation discrimination:
  Learning from labeled and unlabeled data,''
\newblock in {\em NIPS}, 1998, pp. 854--860.

\bibitem{scudder1965probability}
H~Scudder,
\newblock ``Probability of error of some adaptive pattern-recognition
  machines,''
\newblock {\em IEEE Transactions on Information Theory}, vol. 11, no. 3, pp.
  363--371, 1965.

\bibitem{riloff2003learning}
Ellen Riloff, Janyce Wiebe, and Theresa Wilson,
\newblock ``Learning subjective nouns using extraction pattern bootstrapping,''
\newblock in {\em Proceedings of the seventh conference on Natural language
  learning at HLT-NAACL 2003-Volume 4}. Association for Computational
  Linguistics, 2003, pp. 25--32.

\bibitem{zhu2003semi}
Xiaojin Zhu, Zoubin Ghahramani, and John~D Lafferty,
\newblock ``Semi-supervised learning using gaussian fields and harmonic
  functions,''
\newblock in {\em Proceedings of the 20th International conference on Machine
  learning (ICML-03)}, 2003, pp. 912--919.

\bibitem{corduneanu2002information}
Adrian Corduneanu and Tommi Jaakkola,
\newblock ``On information regularization,''
\newblock in {\em Proceedings of the Nineteenth conference on Uncertainty in
  Artificial Intelligence}. Morgan Kaufmann Publishers Inc., 2002, pp.
  151--158.

\bibitem{subramanya2009entropic}
Amarnag Subramanya and Jeff~A Bilmes,
\newblock ``Entropic graph regularization in non-parametric semi-supervised
  classification,''
\newblock in {\em Advances in Neural Information Processing Systems}, 2009, pp.
  1803--1811.

\bibitem{subramanya2008soft}
Amarnag Subramanya and Jeff Bilmes,
\newblock ``Soft-supervised learning for text classification,''
\newblock in {\em Proceedings of the Conference on Empirical Methods in Natural
  Language Processing}. Association for Computational Linguistics, 2008, pp.
  1090--1099.

\bibitem{ozaki2011using}
Kohei Ozaki, Masashi Shimbo, Mamoru Komachi, and Yuji Matsumoto,
\newblock ``Using the mutual k-nearest neighbor graphs for semi-supervised
  classification of natural language data,''
\newblock in {\em Proceedings of the fifteenth conference on computational
  natural language learning}. Association for Computational Linguistics, 2011,
  pp. 154--162.

\bibitem{alexandrescu2009graph}
Andrei Alexandrescu and Katrin Kirchhoff,
\newblock ``Graph-based learning for statistical machine translation,''
\newblock in {\em Proceedings of Human Language Technologies: The 2009 Annual
  Conference of the North American Chapter of the Association for Computational
  Linguistics}. Association for Computational Linguistics, 2009, pp. 119--127.

\bibitem{subramanya2010efficient}
Amarnag Subramanya, Slav Petrov, and Fernando Pereira,
\newblock ``Efficient graph-based semi-supervised learning of structured
  tagging models,''
\newblock in {\em Proceedings of the 2010 Conference on Empirical Methods in
  Natural Language Processing}. Association for Computational Linguistics,
  2010, pp. 167--176.

\bibitem{hakimov2012named}
Sherzod Hakimov, Salih~Atilay Oto, and Erdogan Dogdu,
\newblock ``Named entity recognition and disambiguation using linked data and
  graph-based centrality scoring,''
\newblock in {\em Proceedings of the 4th international workshop on semantic web
  information management}. ACM, 2012, p.~4.

\bibitem{aliannejadi2017graph}
Mohammad Aliannejadi, Masoud Kiaeeha, Shahram Khadivi, and Saeed~Shiry Ghidary,
\newblock ``Graph-based semi-supervised conditional random fields for spoken
  language understanding using unaligned data,''
\newblock {\em arXiv preprint arXiv:1701.08533}, 2017.

\bibitem{mintz2009distant}
Mike Mintz, Steven Bills, Rion Snow, and Dan Jurafsky,
\newblock ``Distant supervision for relation extraction without labeled data,''
\newblock in {\em Proceedings of the Joint Conference of the 47th Annual
  Meeting of the ACL and the 4th International Joint Conference on Natural
  Language Processing of the AFNLP: Volume 2-Volume 2}. Association for
  Computational Linguistics, 2009, pp. 1003--1011.

\bibitem{yao2010collective}
Limin Yao, Sebastian Riedel, and Andrew McCallum,
\newblock ``Collective cross-document relation extraction without labelled
  data,''
\newblock in {\em Proceedings of the 2010 Conference on Empirical Methods in
  Natural Language Processing}. Association for Computational Linguistics,
  2010, pp. 1013--1023.

\bibitem{yao2013universal}
Limin Yao, Sebastian Riedel, and Andrew McCallum,
\newblock ``Universal schema for entity type prediction,''
\newblock in {\em Proceedings of the 2013 workshop on Automated knowledge base
  construction}. ACM, 2013, pp. 79--84.

\bibitem{luan2014relating}
Yi~Luan, Richard Wright, Mari Ostendorf, and Gina-Anne Levow,
\newblock ``Relating automatic vowel space estimates to talker
  intelligibility,''
\newblock in {\em Proc.\ Conf. Int. Speech Communication Assoc. (INTERSPEECH)},
  2014.

\bibitem{levow2014recognition}
Gina-Anne Levow, Valerie Freeman, Alena Hrynkevich, Mari Ostendorf, Richard
  Wright, Julian Chan, Yi~Luan, and Trang Tran,
\newblock ``Recognition of stance strength and polarity in spontaneous
  speech,''
\newblock in {\em Proc.\ IEEE Workshop on Spoken Language Technology (SLT)},
  2014, pp. 236--241.

\bibitem{luan2012performance}
Yi~Luan, Masayuki Suzuki, Yutaka Yamauchi, Nobuaki Minematsu, Shuhei Kato, and
  Keikichi Hirose,
\newblock ``Performance improvement of automatic pronunciation assessment in a
  noisy classroom,''
\newblock in {\em Proc.\ IEEE Workshop on Spoken Language Technology (SLT)}.
  IEEE, 2012, pp. 428--431.

\bibitem{takamatsu2012reducing}
Shingo Takamatsu, Issei Sato, and Hiroshi Nakagawa,
\newblock ``Reducing wrong labels in distant supervision for relation
  extraction,''
\newblock in {\em Proceedings of the 50th Annual Meeting of the Association for
  Computational Linguistics: Long Papers-Volume 1}. Association for
  Computational Linguistics, 2012, pp. 721--729.

\bibitem{zeng2015distant}
Daojian Zeng, Kang Liu, Yubo Chen, and Jun Zhao,
\newblock ``Distant supervision for relation extraction via piecewise
  convolutional neural networks.,''
\newblock in {\em EMNLP}, 2015, pp. 1753--1762.

\bibitem{scienceIE}
Sebastian Riedel Lakshmi Vikraman Andrew~McCallum Isabelle~Augenstein,
  Mrinal~Das,
\newblock ``Semeval 2017 task 10: Scienceie - extracting keyphrases and
  relations from scientific publications,''
\newblock {\em arXiv preprint arXiv:1704.02853}, 2017.

\bibitem{liu2014graph}
Yuzong Liu and Katrin Kirchhoff,
\newblock ``Graph-based semi-supervised acoustic modeling in dnn-based speech
  recognition,''
\newblock in {\em Spoken Language Technology Workshop (SLT), 2014 IEEE}. IEEE,
  2014, pp. 177--182.

\bibitem{subramanya2011semi}
Amarnag Subramanya and Jeff Bilmes,
\newblock ``Semi-supervised learning with measure propagation,''
\newblock {\em Journal of Machine Learning Research}, vol. 12, no. Nov, pp.
  3311--3370, 2011.

\bibitem{aliannejadi2014graph}
Mohammad Aliannejadi, Masoud Kiaeeha, Shahram Khadivi, and Saeed~Shiry Ghidary,
\newblock ``Graph-based semi-supervised conditional random fields for spoken
  language understanding using unaligned data,''
\newblock in {\em Australasian Language Technology Association Workshop}, 2014,
  p.~98.

\bibitem{kim2015weakly}
Young-Bum Kim, Minwoo Jeong, Karl Stratos, and Ruhi Sarikaya,
\newblock ``Weakly supervised slot tagging with partially labeled sequences
  from web search click logs.,''
\newblock in {\em HLT-NAACL}, 2015, pp. 84--92.

\bibitem{luan2017scienceie}
Yi~Luan, Mari Ostendorf, and Hannaneh Hajishirzi,
\newblock ``Scientific information extraction with semi-supervised neural
  tagging,''
\newblock in {\em Proc.\ Conf. Empirical Methods Natural Language Process.
  (EMNLP)}, 2017.

\bibitem{radev2009acl}
Dragomir~R Radev, Pradeep Muthukrishnan, and Vahed Qazvinian,
\newblock ``The acl anthology network corpus,''
\newblock in {\em Proceedings of the 2009 Workshop on Text and Citation
  Analysis for Scholarly Digital Libraries}. Association for Computational
  Linguistics, 2009, pp. 54--61.

\bibitem{bird2008acl}
Steven Bird, Robert Dale, Bonnie~J Dorr, Bryan~R Gibson, Mark~Thomas Joseph,
  Min-Yen Kan, Dongwon Lee, Brett Powley, Dragomir~R Radev, Yee~Fan Tan,
  et~al.,
\newblock ``The acl anthology reference corpus: A reference dataset for
  bibliographic research in computational linguistics.,''
\newblock in {\em LREC}, 2008.

\bibitem{councill2008parscit}
Isaac~G Councill, C~Lee Giles, and Min-Yen Kan,
\newblock ``Parscit: an open-source crf reference string parsing package.,''
\newblock in {\em LREC}, 2008, vol. 2008.

\bibitem{luong2012logical}
Minh-Thang Luong, Thuy~Dung Nguyen, and Min-Yen Kan,
\newblock ``Logical structure recovery in scholarly articles with rich document
  features,''
\newblock {\em Multimedia Storage and Retrieval Innovations for Digital Library
  Systems}, vol. 270, 2012.

\bibitem{ling2015design}
Xiao Ling, Sameer Singh, and Daniel~S Weld,
\newblock ``Design challenges for entity linking,''
\newblock {\em Transactions of the Association for Computational Linguistics},
  vol. 3, pp. 315--328, 2015.

\bibitem{han2011generative}
Xianpei Han and Le~Sun,
\newblock ``A generative entity-mention model for linking entities with
  knowledge base,''
\newblock in {\em Proceedings of the 49th Annual Meeting of the Association for
  Computational Linguistics: Human Language Technologies-Volume 1}. Association
  for Computational Linguistics, 2011, pp. 945--954.

\bibitem{trouillon2017knowledge}
Th{\'e}o Trouillon, Christopher~R Dance, Johannes Welbl, Sebastian Riedel,
  {\'E}ric Gaussier, and Guillaume Bouchard,
\newblock ``Knowledge graph completion via complex tensor factorization,''
\newblock {\em arXiv preprint arXiv:1702.06879}, 2017.

\bibitem{schutze2008introduction}
Hinrich Sch{\"u}tze,
\newblock ``Introduction to information retrieval,''
\newblock in {\em Proceedings of the international communication of association
  for computing machinery conference}, 2008.

\end{thebibliography}

\end{document}